\documentclass[10pt]{article} 
\usepackage{amsmath} 
\usepackage{graphicx}
\usepackage{bm}
\usepackage{cite}
\usepackage{algorithm} 
\usepackage{algpseudocode}
\usepackage{pgfplots}
\usepackage{subfigure}
\usepackage{amssymb}
\usepackage[colorlinks,bookmarksopen,bookmarksnumbered,citecolor=red,urlcolor=red]{hyperref}
\newcommand{\dt}{\Delta t}

\newcommand{\textsub}[2]{{#1}_{\text{#2}}}

\makeatother

\newcommand{\T}{{\text{T}}}
\newcommand{\tv}[1]{\textbf{#1}}

\newcommand{\tm}[1]{\textbf{#1}}

\newcommand{\tran}[1]{^{\text{#1}}}
\newcommand{\bg}[1]{\boldsymbol#1}
\newcommand{\ubar}[1]{\text{\b{$#1$}}}
\newcommand{\RN}[1]{\textup{\uppercase\expandafter{\romannumeral#1}}}

\begin{document}

\title{Adaptive model reduction for nonsmooth discrete element simulation}
\author{Martin Servin \and Da Wang}

\maketitle


\begin{abstract}
A method for adaptive model order reduction for nonsmooth discrete element 
simulation is developed and analysed in numerical experiments.  
Regions of the granular media that collectively move as rigid bodies are
substituted with rigid bodies of the corresponding shape and 
mass distribution.  The method also support particles merging with articulated
multibody systems.  A model approximation error is defined and used to derive conditions for when and where to apply reduction and refinement back into particles and smaller rigid bodies.
Three methods for refinement are proposed and tested: prediction from contact events, 
trial solutions computed in the background
and using split sensors.  The computational performance can be increased by
5 - 50 times for model reduction level between 70 - 95 \%.
\end{abstract}

\section{Introduction}
\label{sec:introduction}

Simulation of granular matter is important for increased understanding
of the nature of granular media and as an engineering tool for design, control
and optimization of processing and transportation systems \cite{poeschel:2005:cgd}.  With the discrete element method (DEM) the material is modeled as a system of contacting rigid 
bodies, referred to as particles in this text.  This provides detailed 
information about force structures and particle kinematics
on a microscopic level.  DEM accurately capture many of the characteristic
phenomena of granular media \-- for instance jamming, dilatancy, 
emergence of strong force chains, strain localization, avalanches and size-segregation 
upon fluidization \-- that are difficult or even impossible to model with continuum based
methods.  The required computational time increase with the number of bodies
and this limit the practical use of DEM for exhaustive simulation studies of 
large-scale systems and large parameter spaces.  One strategy to remedy this
is to increase the computational performance by use of parallel algorithms and
dedicated hardware.  Another strategy is the use of implicit integration with large time-step using the 
nonsmooth DEM (NDEM) \cite{Radjai:2009:cdn}, also known as the contact dynamics method \cite{Moreau:1999:NAS,Jean:1999:NSC},
where velocities may be time-discontinuous and impulses can
propagate instantly through the system.

A third strategy, that is pursued in the paper, is to reduce the computational 
complexity by identifying regions in the granular media where the particles
may be substituted by approximate models with less degrees of freedom.

\subsection{Previous work}
Model order reduction is well established and widely used for reducing the computational 
complexity in solid and fluid mechanics, dynamical systems
and control theory \cite{Kerschen:2005:mpo,Antoulas:2005:als}.  In multibody
dynamics, it is often used for reducing the degrees of freedom of flexible 
bodies \cite{Nowakowski:2012:mor} while the number of multibodies are preserved.
There are few examples that resemble model order reduction for the discrete element method.
Gl\"{o}ssmann \cite{Glossman:2010:rde} applied the Karhunen-Lo\'{e}ve transform to clusters of discrete elements that
show dynamic coherence to reduce the order of generalized coordinates.
In the combined finite-discrete element method (FDEM) each body is
represented as a discrete element that is also 
discretized by a finite element method \cite{Munjiza:2004:cfd}.  The bodies may
deform, fracture and fragment indefinitely into smaller elements based on the
internal stresses.  An inversion of this
is the hierarchical multiscale modeling of granular media 
discretized by a coarse mesh of finite elements in combination with
assemblies of fine grained discrete elements for numerical
computation of the local constitutive law for the finite element computations
\cite{Miehe:2010:hts,Guo:2014:cfd}.  A two-scale and two-method approach 
for modeling granular materials is presented in \cite{Wellman:2012:cfd}, where
DEM is used for domains of large and discontinuous deformations and as an elastoplastic
solid using FEM in continuous domains.  Automatic simplification algorithms of 
articulated multibody systems have been developed and shown to increase
computational performance by two orders in magnitude on large-scale 
linkage systems \cite{Redon:2005:eea,Redon:2005:ada}.  
Dynamic formation of both rigid aggregates (clumped particles) and
elastic aggregates (clustered particles), are supported by several
discrete element codes and is used for modeling grains in brittle rock 
\cite{Cho:2007:rcpm}.

\subsection{Outline of the idea and the challenges}
The idea is to identify regions of the granular
media that collectively move as rigid bodies and substitute each of these
regions with rigid aggregates of the corresponding shape and mass distribution.
Particles and rigid aggregates may also merge with rigid bodies or kinematic
geometries that do not represent granular media, for instance the particles
in an excavator bucket may merge with the bucket into one single rigid body.
The aggregated rigid bodies still contribute to the system dynamics 
but require only a few degrees of freedom.  When merged material is disturbed, 
by a change in external forces or boundary contacts, it may split into smaller constituents that 
are either rigid aggregates of fewer particles or single particles.

The complexity of systems with granular media in the solid state is thus
largely reduced and the computational performance increase correspondingly
while the macroscopic dynamics may be preserved.  For granular media in the gaseous or liquid 
state, on the other hand, a model reduction into rigid aggregates will cause significant errors.
This paper is limited to model reduction into rigid bodies.  The extension to elastoplastic
bodies can be imagined but is beyond the current scope.  The ultimate goal is to achieve optimal 
trade-off between: maximum system reduction; minimum errors on the macroscopic dynamics; 
minimum computational overhead. 

The main challenge is to predict when and where merged material should be refined, or split.
Most real systems are in a combination of the three states of granular media: 
solid, fluid and gas \cite{Jaeger:1996:gsl}.
If the split conditions are too restrictive, or the merge condition
too progressive, the bulk properties become wrong.  This may appear
as incorrect angle of repose, artificial resistance to compression
and shearing forces and erroneous rheology in the fluid state.
If the split conditions are too permissive, or the merge conditions too strict,
most particle will remain free and the there is no computational gain.  Also,
the computational overhead of model reduction must be
small in comparison to the computational time for the fully resolved system.

\begin{figure}
\centering
\includegraphics[width=0.5\textwidth, trim = 0 0 0 0, clip]{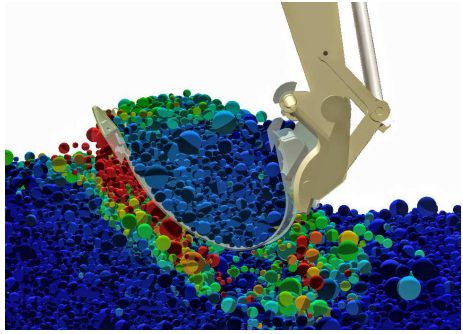}
  \caption{Illustration of model reduction in an excavation scenario simulated without model reduction.  The particles are color coded
  by velocity from blue (stationary) to red.  Most particles in the bed are in relative rest and would be 
  well described by a single rigid body.  The particles in the excavator bucket may
  be aggregated with the bucket into a single rigid body.  The main challenge is to predict
  when and where the merged material should split.}
  \label{fig:illustration}
\end{figure}

The idea is illustrated in Fig.~\ref{fig:illustration} with
an excavator digging in a bed of granular material.  Only a finite domain of
the material around the bucket is displaced and need to be simulated dynamically.
The remaining part is static and contribute merely with supporting contact pressure.
When the bucket is filled and starts to lift, most material co-move rigidly with the bucket.
If the purpose of the simulation is to compute the dynamics of the excavator and the load
forces in the mechanism, the material and the bucket can be approximated by a single
rigid body.  When the bucket accelerates or rotate slowly the force distribution in
the granular material change and it might start to flow.  Several methods for predicting 
splitting of the rigid aggregates are proposed and tested in numerical experiments.
The method are based either on contact events or estimating force distribution
or particle motion by computations in the background.

\section{Particles, rigid aggregate and multibodies}
\label{sec:representation}

This section is devoted to the mathematical representation of contacting
particles and rigid bodies as multibody systems with nonsmooth dynamics and kinematic
constraints.  

\subsection{Global system variables}
The variables for particles, rigid aggregates and other rigid bodies are components of the 
\emph{global system variables} that we denote $\tv{x}, \tv{v}, \bg{f}$ and $\tv{M}$ and
refer to as generalized position, velocity, force and mass although they are concatenations of linear
and rotational degrees of freedom. Quaternions are used for representing orientations.  
The matrix dimension of the global quantities are 
$\dim(\tv{x}) = 7 N_{\text b}$, $\dim(\tv{v}) = \dim(\tv{f}) = 6 N_{\text b}$, 
$\dim(\tm{M}) = 6 N_{\text b} \times 6 N_{\text b} $, where
$\textsub{N}{b}$ is the total number of bodies.

\subsection{Nonsmooth multibody dynamics}
\label{sec:nmd}
The following equations of motion for nonsmooth multibody dynamics are assumed:
\begin{eqnarray}
	\tm{M} \dot{\tv{v}} + \dot{\tm{M}} \tv{v} & = & \textsub{\tv{f}}{ext} + \textsub{\tm{G}}{n}^\T \textsub{\bg{\lambda}}{n} + \textsub{\tm{G}}{t}^\T \textsub{\bg{\lambda}}{t} +  \textsub{\tm{G}}{r}^\T \textsub{\bg{\lambda}}{r} + \textsub{\tm{G}}{j}^\T \textsub{\bg{\lambda}}{j} \label{eq:newton_euler},\\
	0 \leq \textsub{\varepsilon}{n}\textsub{\bg{\lambda}}{n} + \textsub{\tm{g}}{n} + \textsub{\tau}{n}\textsub{\tm{G}}{n}\tm{v}& \perp & \textsub{\bg{\lambda}}{n} \geq 0\label{eq:normal_constraint_reg},\\
	\textsub{\gamma}{t}\textsub{\bg{\lambda}}{t} + \textsub{\tm{G}}{t}\tm{v} & = & 0 ,\ \ 
	|\textsub{\bg{\lambda}}{t}^{(\alpha)}| \leq 
	\textsub{\mu}{t}|\tm{G}_{\text{n}}^{(\alpha)\T}\textsub{\lambda}{n}^{(\alpha)}|\label{eq:coulomb},\\
 	\textsub{\gamma}{r}\textsub{\bg{\lambda}}{r} + \textsub{\tm{G}}{r}\tm{v} & = & 0 \label{eq:rolling},\ \ 
	|\textsub{\bg{\lambda}}{r}^{(\alpha)}| \leq 
	\textsub{\mu}{r}r|\tm{G}_{\text{n}}^{(\alpha)\T}\textsub{\lambda}{n}^{(\alpha)}|,\\
	\textsub{\varepsilon}{j} \textsub{\bg{\lambda}}{j} + \textsub{\eta}{j}\textsub{\tm{g}}{j} + \textsub{\tau}{j}\textsub{\tm{G}}{j}\tm{v} & = & 0. \label{eq:constraint_reg}
\end{eqnarray}
The first equation is the Newton-Euler equation of motion for rigid bodies with external (smooth) 
forces $\textsub{\tv{f}}{ext}$ and constraint force $\textsub{\tm{G}}{}^\T\bg{\lambda}_{}$
with Lagrange multiplier $\bg{\lambda}$ and Jacobian $\tm{G}$, divided into normal (n), tangential (t), 
rolling (r) and articulated and possibly motorized joints (j).  Details are found in the Appendix.  Equations (\ref{eq:normal_constraint_reg})-(\ref{eq:coulomb})
are the Signorini-Coulomb conditions with constraint regularization and 
stabilization terms $\textsub{\varepsilon}{n}$, $\textsub{\tau}{n}$
and $\textsub{\gamma}{t}$.  With $\textsub{\varepsilon}{n} = \textsub{\tau}{n} = 0$,
Eq.~(\ref{eq:normal_constraint_reg}) state that bodies should be separated or 
have zero overlap, $\textsub{\tm{g}}{n}(\tm{x}) \geq 0$, and if so the normal force should be non-cohesive, $\textsub{\bg{\lambda}}{n} \geq 0$.  With $\textsub{\gamma}{t} = 0$, Eq.~(\ref{eq:coulomb}) state
that contacts should have zero relative slide velocity, $\textsub{\tm{G}}{t}\tm{v} = 0$, 
provided that the friction force remain bounded by the Coulomb friction law with friction
coefficient $\textsub{\mu}{t}$.  Eq.~(\ref{eq:rolling}) similarly constrains
relative rotation of contacting bodies provided the constraint torque do not
exceed the rolling resistance law with rolling resistance coefficient 
$\textsub{\mu}{r}$ and radius $r$.  
The constraint force, $\textsub{\tm{G}}{j}^\T \textsub{\bg{\lambda}}{j}$, 
arise for articulated rigid bodies jointed with kinematic links and motors represented 
with the generic constraint equation (\ref{eq:constraint_reg}).  With $\textsub{\varepsilon}{j},
\textsub{\tau}{j} = 0$ and $\textsub{\eta}{j} = 1$, it
become an ideal holonomic constraint $\tm{g}(\tm{x}) = 0$.  For $\varepsilon,\eta = 0$ and $\tau = 1$,
it become an ideal Pfaffian constraint $\tm{G}\dot{\tm{x}} = 0$.  With $\varepsilon, \eta, \tau \neq 0$
it can represent a generic constraint with compliance and damping.  
The set of equations (\ref{eq:newton_euler})-(\ref{eq:constraint_reg}) may thus model granular materials strongly coupled with mechatronic systems,
e.g., vehicles, robots and mechanical processing units.

The Lagrange multiplier become an auxiliary variable to solve for in addition to position and velocity.
The regularization and stabilization terms, $\varepsilon$
and $\gamma$, introduce compliance and dissipation in motion orthogonal
to the constraint manifold.  In the absence of the inequality and complementarity 
conditions, the regularized constraints may be viewed as Legendre transforms 
of a potential and Rayleigh dissipation function of the form 
$U_{\varepsilon}(\tv{x}) = \tfrac{1}{2\varepsilon}\tv{g}^T\tv{g}$
and $\mathcal{R}_\gamma(\tv{x},\tv{v}) = \tfrac{1}{2\gamma}(\tm{G}\tv{v})^T(\tm{G}\tv{v})$
\cite{bornemann:1997:hhs,lacoursiere:2007:rvs}.  This enable modeling of arbitrarily
stiff elastic and viscous interactions in terms of constraint forces with direct mapping
between the regularization and stabilization terms to physical material parameters.
This is applied to map the stiffness and damping terms from the nonlinear Hertz contact 
law, or linear spring and dashpot, from conventional (smooth) discrete element
method to the constraint based and nonsmooth discrete element method.  The detailed constraints,
Jacobians and parameters are found in Ref.~\cite{servin:2014:esn} and 
summarized in Appendix A together with the numerical integration scheme used in
this paper.

The dynamics is allowed to be nonsmooth which means that velocities may 
change discontinuously in time. Impacts and frictional stick-slip transitions 
may thus be considered as instantaneous events and propagate immediately 
through the entire system by an impulse transfer altering the velocities from $\tv{v}_{-}$
to $\tv{v}_{+}$.  The contacts are divided into impacts and 
continuous contacts, depending on the magnitude of the incoming relative 
normal velocities $\textsub{\tm{G}}{n} \tv{v}_{-}$.  The impulse transfer
through the system should satisfy the Newton impact law, 
$\textsub{\tm{G}}{n}^{(n)} \tv{v}_{+} = - e\textsub{\tm{G}}{n}^{(n)} \tv{v}_{-}$,
with coefficient of restitution $e$ for the impacts $(n)$,
as well as preserve all remaining constraints $(m)$ on velocity level, $\tm{G}^{(m)}\tv{v}_{+} = 0$.

\subsection{Particles}
\label{sec:particles}
Each elementary granule is referred to as a \emph{particle} and is modeled
as a rigid body with solid geometry.  A particle is either free or part of a 
rigid aggregate of particles.  Each free particle
is represented by a \emph{dynamic} discrete element obeying the equation of 
motions in Sec.~\ref{sec:nmd} and may interact with other particles and rigid
aggregate bodies via contacts.  Each particle that is part of a rigid aggregate
is a \emph{kinematic} discrete element that co-move rigidly with the aggregate body.
No forces are applied to aggregated particles.
For simplicity, particles are assumed spherical but can easily be extended 
to more general geometric shapes. Reference to a specific particle is made by 
latin indices, e.g., $a,b,\hdots = 1,2,\hdots,N_{\text p}$,
and we use square brackets to emphasize particle index.  We use the notations 
$\vec{\bm{x}}_{[a]}, \vec{\bm{v}}_{[a]}, \vec{\bm{f}}_{[a]}$, $m_{[a]}$ and $d_{[a]}$ for position, 
translational velocity, force, mass and diameter, and $\vec{\bm{e}}_{[a]}, \vec{\bm{\omega}}_{[a]}, 
\vec{\bm{\tau}}_{[a]}$ and $\bm{I}_{[a]}$, for orientation, angular velocity, torque and 
inertia tensor.  Spatial components of a vector or matrix is indexed by 
$\alpha,\beta = 1,2,3$ referring to the $x,y,z$ axes in a global Cartesian coordinate system, e.g., $x^{[a]}_{\alpha}$ and $I_{\alpha\beta}$.  We concatenate these variables 
into a particle's generalized position, velocity, force and mass,
denoted $\tv{x}_{[a]}, \tv{v}_{[a]}, \tv{f}_{[a]}$ and $\tv{M}_{[a]}$, with
$\tv{v}_{[a]} = (\vec{\bm{v}}_{[a]}^\T, \vec{\bm{\omega}}_{[a]}^\T)^\T$ etc.\ and
$\tv{M}_{[a]} = \text{diag} (m_{[a]}\bm{1}_{3\times 3},\bm{I}_{[a]})$.

\subsection{Rigid aggregates}
\label{sec:aggregates}
\begin{figure}
\centering
\includegraphics[width=0.35\textwidth]{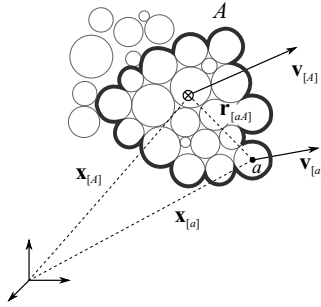}
  \caption{Illustration of a rigid aggregate of contacting particles and some free particles.}
  \label{fig:aggregate}
\end{figure}

A \emph{rigid aggregate}, or just aggregate, is a rigid body that
represent an aggregate of particles co-moving as a single rigid body.
The rigidity is an assumed collective result of the contact forces creating 
a jammed state although no such internal forces are modeled or computed explicitly.
The variables of rigid aggregates are represented with similar notations as for 
particles, e.g., $\tv{x}_{[A]}$ and $\tv{M}_{[A]}$, but with capital latin indices
$A,B,\hdots = 1,2,\hdots,\textsub{N}{a}$.  The set of particles that constitute an aggregate 
$A$ is denoted $\mathcal{N}_A$.  The relation between the dynamic variables of the aggregate and 
the particles are illustrated in Fig.~\ref{fig:aggregate} and computed as follows:  
\begin{eqnarray}
	m_{[A]} & = & \sum_{a\in \mathcal{N}_A} m_{[a]} \label{eq:agglomerate_mass},\\
	\vec{\bm{x}}_{[A]} & = & m_{[A]}^{-1}\sum_{a\in \mathcal{N}_A} m_{[a]}\vec{\bm{x}}_{[a]},\\
	\vec{\bm{v}}_{[A]} & = & m_{[A]}^{-1}\sum_{a\in \mathcal{N}_A} m_{[a]}\vec{\bm{v}}_{[a]},\\
	I^{[A]}_{\alpha\beta}  & = & \sum_{a\in \mathcal{N}_A} m_{[a]}\left( |\vec{\bm{r}}_{[aA]}|^2\delta_{\alpha\beta} - r^{[aA]}_{\alpha} r^{[aA]}_{\beta}\right),\\
	\vec{\bm{\omega}}_{[A]} & = & \bm{I}^{-1}_{[A]} \sum_{a\in \mathcal{N}_A} m_{[a]}\vec{\bm{r}}_{[aA]}\times\vec{\bm{v}}_{[aA]}\label{eq:agglomerate_omega},
\end{eqnarray}
where $\vec{\bm{r}}_{[aA]} = \vec{\bm{x}}_{[a]} - \vec{\bm{x}}_{[A]}$ and 
$\vec{\bm{v}}_{[aA]} = \vec{\bm{v}}_{[a]} - \vec{\bm{v}}_{[A]}$. 
The kinematics of the aggregated particles $a\in \mathcal{N}_A$ is 
\begin{eqnarray}
	\vec{\bm{x}}_{[a]} & = & \vec{\bm{x}}_{[A]} + \vec{\bm{r}}_{[aA]}\label{eq:agg_particle_pos},\\
	\vec{\bm{v}}_{[a]} & = & \vec{\bm{v}}_{[A]} + \vec{\bm{\omega}}_{[A]}\times \vec{\bm{r}}_{[aA]},\\
	\vec{\bm{\omega}}_{[a]} & = & \vec{\bm{\omega}}_{[A]}\label{eq:agg_particle_omega}.
\end{eqnarray}

\subsection{Multibodies}
By elementary rigid bodies it is meant bodies that represent other than granular bodies,
e.g., part of an articulated multibody.  Elementary rigid bodies are also
included in model reduction and may form reduced aggregates with both elementary rigid bodies
and particles.  But the additional complexity of merge and split conditions in articulated system
is not covered here.  For convenience elementary rigid bodies are represented with the same notation as
used for aggregates, that is, capital index in square brackets $[A]$. 

\subsection{Contacts}
The set of contacts is denoted $\textsub{\mathcal{N}}{c}$. Integer $n = 1,2,\hdots,\textsub{N}{c}$ 
is used for contact index and this is emphasized by round 
brackets, e.g., $g^{(n)}$.  The gap function $\delta(\tv{x})$ measure the 
magnitude of overlap between two contacting bodies.  
Contact forces and velocities are sometimes decomposed in 
the directions of contact normal, $\vec{\tv{n}}$ and tangents, $\vec{\mathbf{t}}_1$ and $\vec{\mathbf{t}}_2$.  The relative velocity
at a contact $n$ between a particle $a$ and a rigid body $A$ can thus be written
$\vec{\bm{u}}^{(n)} = \vec{\bm{v}_{[a]}} + \vec{\bm{\omega}}_{[a]}\times\vec{\bm{d}}^{(n)}_{[a]}
- \vec{\bm{v}}_{[A]} - \vec{\bm{\omega}}_{[A]}\times \vec{ \bm{d}}^{(n)}_{[A]}$,
where $\vec{\bm{d}}^{(n)}_{[a]}$ is the position of the contact point relative 
to $\vec{\bm{x}}_{[a]}$ and $\vec{\bm{d}}^{(n)}_{[A]}$ relative to $\vec{\bm{x}}_{[A]}$.

\section{Adaptive model order reduction}
\label{sec:model_reduction}

Let the following equations represent the full system of particles and elementary
rigid bodies coupled with constraints
\begin{eqnarray}
	\tm{M} \ddot{\tv{x}} + \dot{\tm{M}} \dot{\tv{x}}  & = &  \textsub{\tv{f}}{ext} + \tm{G}^\T \bg{\lambda}\label{eq:newton_euler_original},\\
	\varepsilon \bg{\lambda} + \eta\tm{g}(\tm{x}) + \tau\tm{G}\dot{\tm{x}} & = & 0, \label{eq:constraint_reg_original}
\end{eqnarray}
having solution $\tv{x}\in\mathbb{R}^{n}$ and $\bg{\lambda}\in\mathbb{R}^{m}$.  The constraint equation 
(\ref{eq:constraint_reg_original}) represent the collection of both position
and velocity constraints and it is assumed to be appended with additional inequalities and
complementarity conditions for the multiplier.  The full system, 
(\ref{eq:newton_euler_original})-(\ref{eq:constraint_reg_original}), is approximated 
with a reduced system 
$\tilde{\tv{x}}\in\mathbb{R}^{\tilde{n}}$ and $\tilde{\bg{\lambda}}\in\mathbb{R}^{\tilde{m}}$
with less degrees of freedom $\tilde{n}<n$ and $\tilde{m} < m$.  The reduced system 
belong to a subspace of the full system.  We define the model order reduction level as
$h = 1 - \tilde{n}/n$.  When $h \to 1$ the system is maximally reduced to one single rigid body
and when $h = 0$ it is fully resolved in all free particles.  The approximate relation between the reduced and 
full system is expressed using subspace transformation matrices $\tm{P}\in 
\mathbb{R}^{n \times \tilde{n}}$ and $\tm{Q}\in \mathbb{R}^{m \times \tilde{m}}$ such that
\begin{eqnarray}
	\tv{x}  & \approx & \tm{P}\tilde{\tv{x}}\label{eq:x_projection},\\
	\bg{\lambda}  & \approx & \tm{Q}\tilde{\bg{\lambda}}\label{eq:lambda_projection}.
\end{eqnarray}
Given a rigid aggregate, the transformation matrix $\tm{P}$ is easily constructed from 
the rigid transformations in Eq.~(\ref{eq:agg_particle_pos}), that relate the positions of the aggregated particles 
relative to the aggregate centre.  In a rigid aggregate the inter-particle 
constraints are redundant.  The transformation matrix $\tm{Q}$ 
eliminate the redundant equations when a rigid aggregate is substituted 
for a collection of particles.  The reduced multibody system become
\begin{eqnarray}
	\tilde{\tm{M}} \ddot{\tilde{\tv{x}}}  & =  & \textsub{\tilde{\tv{f}}}{e} + \tilde{\tm{G}}^\T  \tilde{\bg{\lambda}} \label{eq:newton_euler_reduced},\\
	\varepsilon \tilde{\bg{\lambda}} + \eta\tilde{\tv{g}}(\tilde{\tv{x}}) + \tau\tilde{\tm{G}}\dot{\tilde{\tv{x}}} & = & 0 \label{eq:constraint_reg_reduced}
\end{eqnarray}
with $\tilde{\tm{M}} = \tm{P}\tm{M}\tm{P}^T$,
$\textsub{\tilde{\tv{f}}}{e} = \tm{P}\textsub{\tv{f}}{ext} - \dot{\tilde{\tm{M}}} \dot{\tilde{\tv{x}}} $,
$\tilde{\tm{G}} = \tm{Q}^T\tm{G}\tm{P}$, $\tilde{\bf{\lambda}} = \tm{Q}\bg{\lambda}$,
$\tilde{\tv{g}} = \tm{Q}\tv{g}$.  Note that $\dot{\tilde{\tm{M}}} \dot{\tilde{\tv{x}}}$ is 
included as an explicit force in $\textsub{\tilde{\tv{f}}}{e}$ as is common and
often referred to as the gyroscopic force.
If the complexity of the reduced system 
is much smaller than the original,  $\tilde{n}\ll n$ and $\tilde{m} \ll m$, the 
computational efficiency can increase dramatically. 

The reduced system can, however, expected to deviate more or less from the
original system.  The model reduction may be applied adaptively to
keep the approximation error below a specified error tolerance. 
The approximation error is defined
\begin{equation}
	\mathcal{E}(t)= \tm{x}(t) -  \tm{P}\tilde{\tm{x}}(t) \label{eq:approx_error}
\end{equation}
and can be decomposed in two orthogonal terms $\mathcal{E}(t) = \mathcal{E}_{\perp}(t) + \mathcal{E}_{\parallel}(t)$,
where
\begin{eqnarray}
	\mathcal{E}_{\perp}(t) & = & \left[ \tm{1} - \tm{P}\tm{P}^T \right] \tm{x}(t) \label{eq:orth_error},\\
	\mathcal{E}_{\parallel}(t) & = & \tm{P} \left[ \tm{P}^T\tm{x}(t) - \tilde{\tm{x}}(t). \right]  \label{eq:parallel_error}
\end{eqnarray}
The orthogonal error, $\mathcal{E}_{\perp}$, arise when the trajectory of the full system 
is not strictly within the subspace of the projection and do not move entirely like a single rigid body.
The error parallel to the projection, $\mathcal{E}_{\parallel}$, means that the motion of the reduced 
system behave different  from the original system although they represent equivalent rigid systems, 
if $\mathcal{E}_{\perp} = 0$.
The parallel error can only be computed a posteriori.  This may be 
practically infeasible as it requires solving both the full and reduced 
system and involve explicit projections between the spaces.  
The orthogonal error, on the other hand, can be estimated a priori.  By 
substituting Eq.~(\ref{eq:constraint_reg_original}) into (\ref{eq:newton_euler_original})
and multiplying by $\tm{P}_{\perp} \equiv  \tm{1} - \tm{P}\tm{P}^T$ one obtain an evolution equation 
for the orthogonal approximation error
\begin{equation}
	\ddot{\mathcal{E}}_{\perp} =  \ubar{\tm{M}}^{-1} \left[ \ubar{\tm{f}}_{\text{e}} 
	-  \frac{\eta }{\varepsilon}\ubar{\tm{G}}^T \tm{g}  - \frac{\tau}{\varepsilon} \ubar{\tm{G}}^T\ubar{\tm{G}} \dot{\ubar{\tm{x}}}\right]
 \label{eq:ddot_approx_error}
\end{equation}
where $\ubar{\tm{M}} = \tm{P}_{\perp}\tm{M}\tm{P}_{\perp}^T$, $\ubar{\tm{f}}_{\text{e}} = 
\tm{P}_{\perp}\tm{f}_{\text{ext}} - \dot{\ubar{\tm{M}}}\dot{\ubar{\tm{x}}}$ 
and $\ubar{\tm{G}} = \tm{G}\tm{P}_{\perp}$.  
The rigid aggregate is a good 
approximation only if the error is small.  When the error is large, or growing rapidly, the reduced
approximation should not be applied.  When particles co-move rigidly,
the third term vanishes, $\ubar{\tm{G}} \dot{\ubar{\tm{x}}} = 0$, since
the relative contact velocity is zero.  The first and second term cancel
when force balance occur in the subspace,  
$\ubar{\tm{f}}_{\text{e}} = \tfrac{\eta }{\varepsilon}\ubar{\tm{G}}^T \tm{g}$.
This is equivalent to zero relative acceleration in the contact points, $\ubar{\tm{G}} \ddot{\ubar{\tm{x}}} = 0$.  Observe that $\ubar{\tm{M}}^{-1} \ubar{\tm{f}}_{\text{e}} = 0$ in the case of uniform gravity,  
since this cause no relative acceleration.  We thus identify the following
conditions for a rigid aggregate to be a good approximation 
\begin{eqnarray}
	-\bg{\xi}_{v}^{-}  < \ubar{\tm{G}}^{(n)} \dot{\ubar{\tm{x}}}   < \bg{\xi}_{v}^{+}\label{eq:error_velocity}
\end{eqnarray}
and
\begin{equation}
	\frac{ \left|\ubar{\tm{f}}_{\text{e}[a]} -  \frac{\eta }{\varepsilon}\ubar{\tm{G}}^T_{[a]} \tm{g}\right|}
	{\left|\ubar{\tm{f}}_{\text{e}[a]}\right| + \left|\frac{\eta }{\varepsilon}\ubar{\tm{G}}^T_{[a]} \tm{g}\right|} \leq  \xi_{f}
	\label{eq:force_balance}
\end{equation}
for each contact $n$ and particle $a$ of the aggregate
with upper and lower thresholds for relative contact velocity 
$\xi_{v}^{\pm}$ and force balance bound $ \xi_{f}$.
The inequalities in Eq.~(\ref{eq:error_velocity}) should be
understood component wise for normal, tangent and rolling.
The conditions (\ref{eq:error_velocity}) and (\ref{eq:force_balance})
provide a starting point for adaptive model order reduction and refinement
by merging and splitting particles into and from rigid aggregates.  Identifying 
wether particles should merge is a simple examination of Eq.~(\ref{eq:error_velocity})
and (\ref{eq:force_balance}).  Predicting if and which particle should 
split is non-trivial since it requires some form of estimation of the unknown dynamics 
of the fully resolved system.  

An algorithm for numerical simulation of nonsmooth multibody systems of the form of 
Eqs.~(\ref{eq:newton_euler})-(\ref{eq:constraint_reg}) and with adaptive model reduction
is given in Algorithm \ref{alg:main}.  The algorithm is based on the SPOOK stepper 
\cite{lacoursiere:2007:rvs} using fix timestep, $\dt$, and involve solving a mixed
complementarity problem (MCP)
with matrix $\tm{H}$, vector $\tv{b}$ and regularization and stabilization matrices $\bg{\Sigma}$
and $\bg{\Upsilon}$ that are found in the Appendix.  A popular choice of MCP solver for NDEM is
the projected Gauss-Seidel (PGS) method, which is also listed in the Appendix.
It should be straightforward to modify the algorithm to other
time-integration schemes and solver methods for nonsmooth dynamical systems.  The test for model 
reduction is done directly after the continuous MCP solve when the new velocities are known.
If this instead is placed after the position update, some contacts that fulfil condition (\ref{eq:error_velocity})
may be lost due to infinitesimal geometric separation.  This would make the model reduction 
unnecessarily sensitive to solver truncation errors.  Particles with contacts that fulfil the 
test are merged.  The test for model refinement is done after contact detection
and before solving the impact stage MCP.  This way the rigid aggregates can be 
split before the impact impulses are computed and transferred.  Otherwise the granular matter will
behave overly rigid.  The merge and split processes are described in more detail below
as well as a number of methods for predicting model refinement.

\begin{algorithm}[h!]
 \caption{\texttt{main\_algorithm}}\label{alg:main}
  \begin{algorithmic}[1]
  \State define constants and parameters
  \State initialil state: $(\tv{x}_{0}, \tv{v}_{0})$
  \For{$i = 0,1,2,\hdots,t/\dt-1$} \Comment{time-stepping}
  \State $[\tv{g}, \textsub{\mathcal{N}}{c} ] = \texttt{contact\_detection}(\tv{x}_i,\tv{v}_i)$
  \State  $[ \tv{G}, \bg{\Sigma}, \bg{\Upsilon} ] = \texttt{compute\_contact\_data}( \tv{x}_i, \tv{v}_i, \textsub{\mathcal{N}}{c} )$
  \State $\texttt{split}(\tv{x}_i, \tv{v}_i , \tv{g}, \tv{G}, \textsub{\mathcal{N}}{c})$    \Comment{model refinement}
  \State $\tv{H} = \texttt{compute\_H}(\tv{M}, \tv{G}, \bg{\Sigma})$
  \State $\tv{b}^{-} = \texttt{compute\_b\_}(\tv{G}, \tv{v}_{i}, e)$
  \State $[ \tv{v}^{+}_{i}, \bg{\lambda}^{+}_{i} ] = \texttt{mcp} ( \tv{H}, \tv{b}^{-} )$   \Comment{impact stage MCP}
  \State $\tv{b} = \texttt{compute\_b}(\tv{g}, \tv{G}, \tv{v}^{+}_{i}, \textsub{\tv{f}}{ext},  \bg{\Upsilon})$
  \State $[ \tv{v}_{i+1}, \bg{\lambda}_{i+1} ] = \texttt{mcp} ( \tv{H}, \tv{b} )$   \Comment{continuous stage MCP}
  \State $\tv{v}_{i+1} = \texttt{co-move}(\tv{x}_{i}, \tv{v}_{i+1})$ 		\Comment{update aggregate particles}
  \State $\texttt{merge}(\tv{x}_i, \tv{v}_i , \tv{g}, \tv{G}, \textsub{\mathcal{N}}{c})$  \Comment{model reduction}
  \State $\tv{x}_{i+1}  = \tv{x}_i + \dt \tv{v}_{i+1}$	\Comment{position update}
  \EndFor
  \end{algorithmic}
\end{algorithm}

\subsection{Merge}
\label{sec:merge}
\begin{figure}[!ht]
\centering
\includegraphics[width=0.8\textwidth]{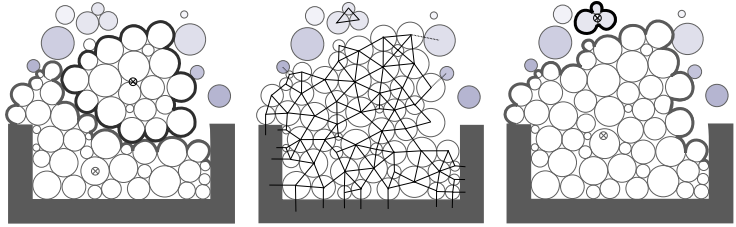}
  \caption{Illustration the merge procedure: detection of contacts involving two aggregates 
  and 11 free particles (left), identification of two disconnected networks fulfilling the merge
  conditions (middle), creating two new aggregates (right).  The colour intensity codes velocity.
    }
  \label{fig:merge}
\end{figure}

The model reduction test consist of traversing the contact network $\textsub{\mathcal{N}}{c}$
and testing the condition for rigid motion in Eq.~(\ref{eq:error_velocity}).
The condition is divided into
\begin{eqnarray}
	\tm{G}^{(n)}_{\text{n}[aA]} \tm{v}  & \in & [- \xi^{\text{i-mrg}}_{\text{n}v}, \xi^{\text{s-mrg}}_{\text{n}v}],\\
	\tm{G}^{(t)}_{\text{t}[aA]} \tm{v}  & \in & [- \xi^{\text{mrg}}_{\text{t}v}, \xi^{\text{mrg}}_{\text{t}v}],\\
	\tm{G}^{(r)}_{\text{r}[aA]} \tm{v}  & \in & [- \xi^{\text{mrg}}_{\text{r}v}, \xi^{\text{mrg}}_{\text{r}v}],
	\label{eq:contact_merge}
\end{eqnarray}
where we separate between incident and separating normal velocity thresholds, $\xi^{\text{i-mrg}}_{\text{n}v}$
and $\xi^{\text{s-mrg}}_{\text{n}v}$, and use symmetric tangential and rolling velocity thresholds $\xi^{\text{mrg}}_{\text{t}v}$
and $\xi^{\text{mrg}}_{\text{r}v}$.  The test result in a set of disconnected networks representing rigidly co-moving bodies. 
Each such network is merged into rigid aggregates.  Both particles and elementary rigid bodies
are allowed to merge into aggregates.  All bodies that are merged into an aggregate body 
are changed from being a dynamic body to a kinematic body co-moving with the new 
aggregate body.  The aggregate variables $m_{[A]}, \vec{\tv{x}}_{[A]}, \vec{\tv{v}}_{[A]},
\vec{\tv{I}}_{[A]}, \vec{\bm{\omega}}_{[A]}$ are computed by Eq.~(\ref{eq:agglomerate_mass})-(\ref{eq:agglomerate_omega}).  
The total mass and momentum is preserved when bodies are merged.
The merge procedure is illustrated in Fig.~\ref{fig:merge}.
It should be emphasized that the contact network need not a force 
network, which would require solving Eq.~(\ref{eq:newton_euler})-(\ref{eq:constraint_reg}), but merely a connectivity network,
and is automatically produced by the contact detection algorithm.
%

\subsection{Split}
\label{sec:split}
\begin{figure}[!ht]
\centering
\includegraphics[width=0.8\textwidth]{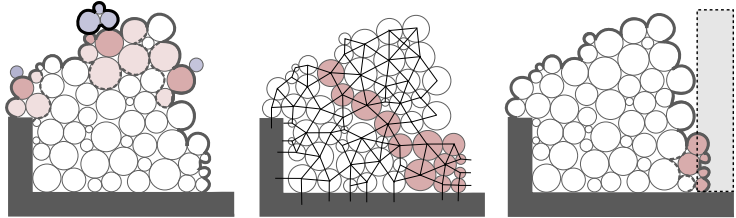}
  \caption{Illustration of different split methods: contact split (left),
  trial solve split (middle) and sensor split (right).  Blue particles have
  impacting or separating contacts. Red particles are aggregate particles that
  are to be split from the aggregate.  The dashed grey box represent a split sensor. }
  \label{fig:split}
\end{figure}

There are a number of ways to predict if and how the reduced model should be refined
by splitting the rigid aggregates into smaller aggregates and free particles.  One strategy is to
rely on contact events, that is, trigger splitting by impacts and separations.  Another strategy
is to do a fast trial solve in the background, using a more resolved model, and decide
splitting from the outcome.  A third, more heuristic approach, is to add split sensors in the
system.  The placement of the sensors can be made automatic, after a posteriori analysis of
previous simulations of the same or similar system, or manually, based on experience or perspicacity.  The split methods can also be used in combination with each other.  The different methods, illustrated in Fig.~\ref{fig:split}, 
are outlined in further detail below and tested in numerical experiments.  
Observe that the split process does not
affect on the total mass or momentum. 
The process of splitting an aggregate is the same irrespective of the method used
for the prediction.  The split particles are made dynamic and the new aggregates
are determined and created as in the merge process, by analysing the contact network
after removing the split particles and applying Eq.~(\ref{eq:agglomerate_mass})-(\ref{eq:agglomerate_omega}).

\subsubsection{Contact split}
\label{sec:impact_split}
Contacts are either impacting, continuous or separating depending on the sign and magnitude
of the relative velocity.  For each contact $n \in \textsub{\mathcal{N}}{c}$, between particle $a$ and aggregate $A$,
the following conditions for normal, tangential and rolling motion are tested
\begin{eqnarray}
	\tm{G}^{(n)}_{\text{n}[aA]} \tm{v}  & \notin & [- \xi^{\text{i-c}}_{\text{n}v}, \xi^{\text{s-c}}_{\text{n}v}],\\
	\tm{G}^{(n)}_{\text{t}[aA]} \tm{v}  & \notin & [- \xi^{\text{c}}_{\text{t}v}, \xi^{\text{c}}_{\text{t}v}],\\
	\tm{G}^{(n)}_{\text{r}[aA]} \tm{v}  & \notin & [- \xi^{\text{c}}_{\text{r}v}, \xi^{\text{c}}_{\text{r}v}],
	\label{eq:contact_split}
\end{eqnarray}
with impact split threshold denoted with \emph{i-c} and separation split threshold 
with \emph{s-c}.  If either of the relative contact velocities are found to be outside the valid domain,
the aggregate is refined by splitting
off particles at the contact node $n$.  The split may be applied to some split depth
$\textsub{N}{c-spl}\in \mathbb{N}$ along the contact network from the impact (or separation) node 
in order to capture shock propagation phenomena.
Observe that the split action is merely a redefinition of which particles are free and
which are kinematically bound to aggregate bodies.  This does not immediately alter the position or 
velocity of any particle.  The split is then proceeded with solving the impact stage MCP and 
the continuous stage MCP, see Algorithm \ref{alg:main}.  There is therefore no
risk, other than unnecessary computations, of splitting too many particles.  These will be merged
back with the aggregate if the merge condition is fulfilled after solving the impact and continuous
MCP.  The numerical experiments presented in this paper is limited to splitting aggregate 
particles that are triggered by impact or separation directly and their contact
neighbours, i.e., $\textsub{N}{c-spl} = 2$.  Impacts between two aggregates are 
treated the same way.

\subsubsection{Trial solve split}
A trial solver is run in the background to estimate the dynamics of the full system.  The background
system state is initialized by  projecting the reduced sub-space system back to the
full resolution space of Eq.~(\ref{eq:agg_particle_pos})-(\ref{eq:agg_particle_omega}).
The purpose of the background solve is not to do precise integration of the particle positions and velocities
but to provide sufficient estimate for if and how to split rigid aggregates.  Assuming that the state of the reduced system
from previous timestep was a good approximation of the full system it is conjectured that doing a
PGS solve of the full system MCP with low number of iterations $N_{\text{it}}^{\text{tr}}$ will suffice for this, although
the error of such a simulation would increase rapidly over time.  Other alternatives can be imagined,
e.g., doing the background computation using other solvers or on a partially resolved system.
The background trial solution is tested for the following conditions for relative velocity of each contact
$n\in\textsub{\mathcal{N}}{c}$ 
\begin{eqnarray}
	\tm{G}^{(n)}_{\text{n}[aA]} \tm{v}  & \notin & [-\xi^{\text{i-tr}}_{\text{n}v}, \xi^{\text{s-tr}}_{\text{n}v}],\\
	\tm{G}^{(n)}_{\text{t}[aA]} \tm{v}  & \notin & [- \xi^{\text{tr}}_{\text{t}v}, \xi^{\text{tr}}_{\text{t}v}],\\
	\tm{G}^{(n)}_{\text{r}[aA]} \tm{v}  & \notin & [- \xi^{\text{tr}}_{\text{r}v}, \xi^{\text{tr}}_{\text{r}v}],
	\label{eq:trial_split_contact}
\end{eqnarray}
and the following conditions for force and torque balance are tested separately for each aggregated particle $a$ 
\begin{eqnarray}
\frac{|\vec{\tm{f}}_{\text{e}[a]} -  \tm{G}^T_{f[a]} \bm{\lambda}|}
	{|\vec{\tm{f}}_{\text{e}[a]}| + |\tm{G}^T_{f[a]} \bm{\lambda}|}  & < & \xi^{\text{tr}}_{f},\\
	\frac{ |\vec{\bm{\tau}}_{\text{e}[a]} -  \tm{G}^T_{\tau[a]} \bm{\lambda}|}
	{|\vec{\bm{\tau}}_{\text{e}[a]}| + |\tm{G}^T_{\tau[a]} \lambda|}  & < & \xi^{\text{tr}}_{\tau}.
	\label{eq:trial_split_particle}
\end{eqnarray}
where $\tm{G}^T_{f[a]} \bm{\lambda}$ and $\tm{G}^T_{\tau[a]} \bm{\lambda}$ are the force and torque
components of the sum of generalized contact forces acting on particle $a$, $\tm{G}^T_{[a]} \bm{\lambda} = \sum_{b\in\mathcal{N}_{\text{c}}^{a}} \tm{G}^T_{[ab]} \bm{\lambda}$.  The Jacobians
are the blocks for linear and rotational degrees of freedom $\tm{G}^T_{[a]} = [ \tm{G}^T_{f[a]}, \tm{G}^T_{\tau[a]}]$.
Observe that $\bm{\lambda}$ has replaced $\tfrac{\eta}{\varepsilon}\tv{g}$ in the force balance
condition (\ref{eq:force_balance}).  This is a stronger test as it can detect also acceleration due
to impulse force propagation through the system that has not yet resulted in relative
contact velocity or particle displacements.  The particles that are indicated by 
the tests are eliminated from the aggregate body and activated as a dynamic particle.  In the numerical 
implementation a small perturbation is added to the denominators
to avoid numerical round-off errors. 
  
\subsubsection{Sensor split}
The split sensor is simply a geometrical shape that triggers model refinement of aggregate bodies
that overlap with the sensor geometry.  The splitting is applied only to the aggregate particles
that overlap the sensor.  Observe that the sensors are physically transparent and do not produce
any contact forces.  Split sensors is a useful tool for when it can be anticipated where
model refinement is required without doing a background trial solve.  The sensor must be given
a size, shape and position, either manually or automated based on data from simulations, models or
experiments.

\section{Numerical experiments}
The described method for adaptive model order reduction is investigated in numerical experiments.
The test systems are a conveyor with a continuous formation of a pile on one end and 
discharge at the other end, 
a granular collapse and granular flow in a slowly rotating drum.  The granular dynamics from using model reduction
is compared with reference simulations run in full resolution.  The achieved model order
reduction level, $h(t) = 1 - \tilde{N}_{\text{p}}(t) / N_{\text{p}}$, is monitored.  
The tests are selected to represent different types of flow and transitions between static and
dynamic states. Model reduction can be expected to work well for the formation of piles where the
dynamics mainly occur in the surface layers. For granular discharge and collapse there is high risk of
approximation errors by to predict when and where the aggregate should split and flow. In a slowly
rotating drum the granular flow separate in one zone of rapid flow (shear zone) on top of
a plug zone that co-rotate with the drum. The plug zone does not rotate as an ideal rigid body,
however, but has a small shear creep \cite{renouf:2005:nst}
that may render model reduction into rigid aggregates
problematic.  The simulations are made
using the simulation software AgX Dynamics \cite{agx} with a prototype implementation
of the adaptive model reduction algorithms in Lua scripts \cite{lua}.  The prototype implementation 
is not optimized for speed and memory and the tests are therefore limited to relatively
small systems ranging between $\textsub{N}{p} = 4-90\cdot 10^3$ particles.  The model and NDEM simulation parameters 
are listed in Table \ref{table:model_parameters} and the adaptive model order reduction  parameters 
in Table \ref{table:reduction_parameters}.  A linear contact model is used, with 
normal stiffness $\textsub{k}{n}$, and may easily be replaced by the nonlinear Hertz contact law.   
Mono-sized spherical particles are used, except in the 
granular collapse where bi-disperse spheres are used.  
Gravity acceleration is $9.81$ m/s$^2$.
Videos from simulations are available 
as supplementary material at \url{http://umit.cs.umu.se/modelreduction/}.

\begin{table}	
  \begin{center}
    \caption{Model and NDEM parameters}\label{table:model_parameters}
    \begin{tabular}{|l|l|}
      \hline
      $d$ 	& 	$13, 10$ mm\\
      $\rho$	&	$3700$ kg/m$^3$\\
      $\textsub{k}{n}$	& 	$3$ kN/m\\
      $e$	& 	$0.18$\\
      $\textsub{\mu}{t}$	& 	$0.91$\\
      $\textsub{\mu}{r}$	& 	$0.32$\\
      $\dt$			&	$5$ ms\\
      $\textsub{N}{p}$	&	$4-90\cdot 10^3$\\
      $\textsub{N}{it}$	&	$150$\\
      $N_\text{it}^\text{ref}$&	$150,500$\\
      \hline
    \end{tabular}
  \end{center}
\end{table}  
\begin{table}	
  \begin{center}
    \caption{Model order reduction parameters}\label{table:reduction_parameters}
    \begin{tabular}{|l|l|}
      \hline
      \bf{parameter} & \bf{value}\\
      \hline
      $\xi_{\text{n}v}^{\text{i-mrg}}$	& 	$2.5$ mm/s\\
      $\xi_{\text{n}v}^{\text{s-mrg}}$	& 	$2.5$ mm/s\\
      $\xi_{\text{t}v}^{\text{mrg}}$	& 	$2.5$ mm/s\\
      $\xi_{\text{r}v}^{\text{mrg}}$	& 	$0.5$ rad/s\\
      $\xi_{\text{n}a}^{\text{i-mrg}}$	& 	$5$ m/s$^2$\\
      $\xi_{\text{n}a}^{\text{s-mrg}}$	& 	$5$ m/s$^2$\\
      $\xi_{\text{t}a}^{\text{mrg}}$	& 	$5$ m/s$^2$\\
      $\xi_{\text{r}a}^{\text{mrg}}$	& 	$\infty$ rad/s\\
       \hline
      $\textsub{N}{c-spl}$ &	$2$\\
      $ \xi^{\text{i-c}}_{\text{n}v}$	& 	$0.15$ m/s\\
      $ \xi^{\text{s-c}}_{\text{n}v}$	& 	$\infty$ m/s\\
      $ \xi^{\text{c}}_{\text{t}v}$	& 	$0.15$ m/s\\
      $ \xi^{\text{c}}_{\text{r}v}$	& 	$\infty$  rad/s\\
            \hline
      $N_\text{it}^\text{tr}$&	$50,100$\\
      $ \xi^{\text{i-tr}}_{\text{n}v}$	& 	$2.6$ m/s\\
      $ \xi^{\text{s-tr}}_{\text{n}v}$	& 	$0.26$ m/s\\
      $ \xi^{\text{tr}}_{\text{t}v}$	& 	$0.26$ m/s\\
      $ \xi^{\text{tr}}_{\text{r}v}$	& 	$\infty$ rad/s\\	
      $ \xi^{\text{tr}}_{f}$			& 	$0.15, 0.25$\\  
      $ \xi^{\text{tr}}_{\tau}$		& 	$\infty$\\
      \hline
    \end{tabular}
  \end{center}
\end{table}  

%

\begin{figure}
\centering
  \includegraphics[width=0.85\textwidth, trim = 0 0 0 0, clip]{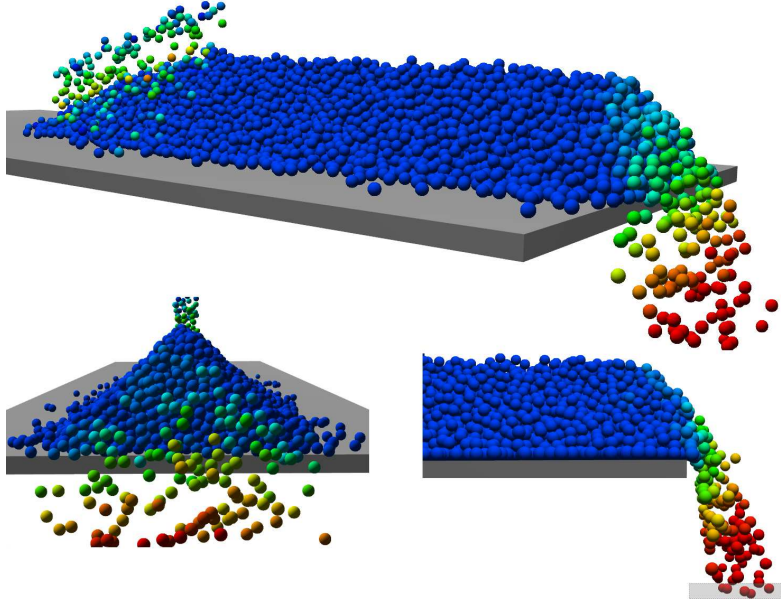}
  \caption{Simulation test for pile formation and discharge on the ends of a conveyor.  The
  colour is coded by particle velocity relative to the conveyor speed and ranges from 
  $0$ (blue) to $1.5$ m/s or above (red).}
  \label{fig:pile_formation}
\end{figure}

The pile formation is performed by emitting particles at a rate of $1000$ s$^{-1}$ from
$0.1$ m above a planar conveyor surface moving with horizontal speed $0.1$ m/s.  The emitter surface
is $15d \times 4 d$, in which the particle positions are chosen randomly.  The particles quickly come to relative rest on the conveyor, forming an elongated pile,
 roughly $10 d$ high and with static angle repose $\theta_{\text{conv}}$.
The angle of repose is computed as the average inclination of the pile surface 
defined by the surface particles in the mid section of the conveyor, neglecting particles resting directly on the conveyor surface, see Fig.~\ref{fig:pile_formation}.  
The discharge take place on the end of the $50 d$ long conveyor, where the material loose support and flow over the edge.  The 
cross sectional flow distribution in the horizontal plane is measured $15 d$ below the conveying surface
 and the geometric centre of the flow is computed, $(\textsub{x}{c},\textsub{y}{c})$.  
 The result is time averaged over $6$ s.  The conveyor
system involves $90\cdot 10^3$ particles when the conveyor is filled.  

\begin{figure}
\centering
  \includegraphics[height=0.9\textwidth]{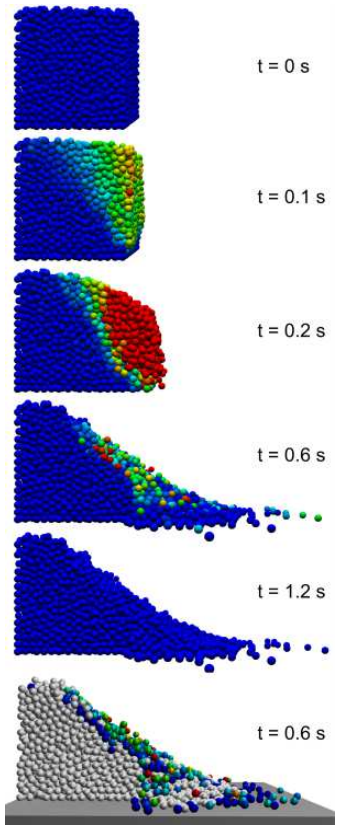}
  \caption{Image sequence from granular collapse.  The
  colour is coded by particle velocity ranging from 
  $0$ (blue) to $0.5$ m/s or above (red).  The top five figures are reference simulation
  and the bottom figure uses model reduction with background trial solve split.}
  \label{fig:dam_break_config}
\end{figure}

The granular collapse
test is made with $4000$ particles emitted randomly from above into a frictionless cubic container with side length $15 d$.  The particles are left to comte to rest
before model reduction is applied.
One side-wall is raised quickly and the particles are left to collapse by the new and unstable force configuration.  Since
the side walls are frictionless the raising of the wall does not disturb the particles other than the change in the confining pressure.  The granular collapse last for about $0.7$ s, 
after which the particle have come to rest  in a semi-pile with well-defined angle of repose, 
$\textsub{\theta}{collapse}$.  The evolution of the angle of repose is tracked
by estimating the motion of the plane defined by the particles on the top surface of the granular cube,
discarding any particles that disconnect from the main contact network.
Images from simulations are shown in Fig.~\ref{fig:dam_break_config}.

\begin{figure}
\centering
  \includegraphics[width=0.4\textwidth, trim = 0 0 0 0, clip]{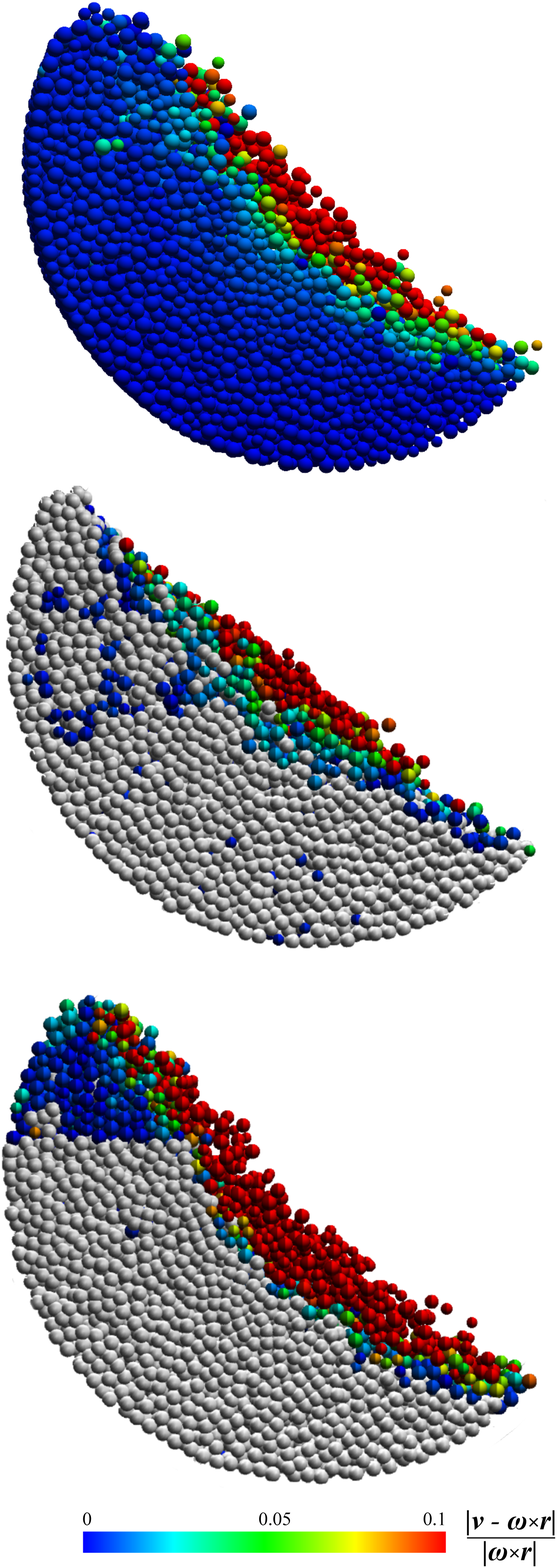}
  \caption{Image from drum flow simulation.  The colour is coded by particle 
  relative velocity to rigid co-motion with the drum, ranging from
  $0$ (blue) to $10\%$ or above (red).  The top figure show the reference simulation,
  the middle one uses model reduction with background trial solve split and the bottom one
  contact event and sensor split.}
  \label{fig:drum_flow}
\end{figure}

The rotating drum has diameter $D = 40 d$, width $w = 7d $ and is run with angular velocity 
$\Omega=0.5$ rad/s.  The corresponding dimensionless Froude number is $\text{Fr} \equiv D\Omega^2/2g \approx 0.01$,
which is in the dense rolling flow regime. 
The side walls are frictionless while the cylinder surface has the sama friction as between particles.
A slow dense nearly stationary flow with $\textsub{N}{p} = 4864$ particles is 
established within one drum revolution without applying model reduction and starting from a regular particle distribution.
An image from the simulations is shown in  Fig.~\ref{fig:drum_flow}.  The particles
have bi-disperse size distribution $d_{1}=13$ mm and $d_{2}=10$ mm.  The
mass distribution and cross sectional flow velocity field are measured and the
dynamic angle of repose, $\theta_{\text{drum}}$, is computed by tracking
the a $D/2$ wide section of the surface around the drum centre.  When using sensor split,
the sensor is placed around the estimated lift height.

\section{Results}

The simulation results are labeled either \emph{ref-500}, \emph{ref-150}, \emph{cs-150}, 
\emph{tr-150-50-f} or \emph{ctr-150-50-fv},
 where the first prefix refer to reference simulation or 
the method for model reduction, the first number is the number or PGS iterations and the
second number is the number iterations in the background trial solver (tr) using force
balance condition only (f) or in conjunction with velocity condition (fv).  
It was found that separation splitting was very sensitive to parameters and typically lead to either a propagation of
splitting over the entire aggregates or not enough splitting.  Therefore it is not applied, i.e.,
$ \xi^{\text{s-c}}_{\text{n}v} = \infty$, and contact splitting need to be combined with either 
sensors (cs) or with trial splitting (ctr).  The results are summarized in Table \ref{table:results}.
\begin{table}	
  \begin{center}
    \caption{Simulation results}\label{table:results}
    \begin{tabular}{|l|l|l|l|l|l|}
      \hline
      			& 	ref-500			&	ref-150		&	cs-150	&	tr-150-50-f&	tr-150-50-fv\\
      \hline
      $\textsub{\theta}{conv}$	&	$41.5\pm0.7^{\circ}$	&	$40.6\pm0.5^{\circ}$&	$40.3\pm0.8^{\circ}$ & & \\
      $h_{\text{conv}}^{\text{mean}}$	& 	$0$ \%		 	&	$0$ \% 		& 	$85$ \% 			& &	\\
      \hline
      $\textsub{\theta}{collapse}$	&	$36^{\circ}$ 	&	$30^{\circ}$ &	& $38^{\circ}$ & $34^{\circ}$\\
     $ h_{\text{collapse}}$	& 	$0$ \%	 &	$0$ \% 		& 	 & $50 - 100$ \% & $50 - 100$ \%	\\
      \hline
      $\textsub{\theta}{drum}$	&	$44\pm3^{\circ}$		& $43\pm2^{\circ}$	& $49\pm5^{\circ}$ & $44\pm3^{\circ}$	 & $45\pm2^{\circ}$	\\
     $h_{\text{drum}}^{\text{mean}}$	& 	$0$ \%		&	$0$ \% 		& $70\pm10$ \%	 & $40\pm10$ \%	 & $40\pm10$ \%	\\
      \hline
    \end{tabular}
  \end{center}
\end{table}  

\begin{figure}
\centering
  \includegraphics[width=0.95\textwidth]{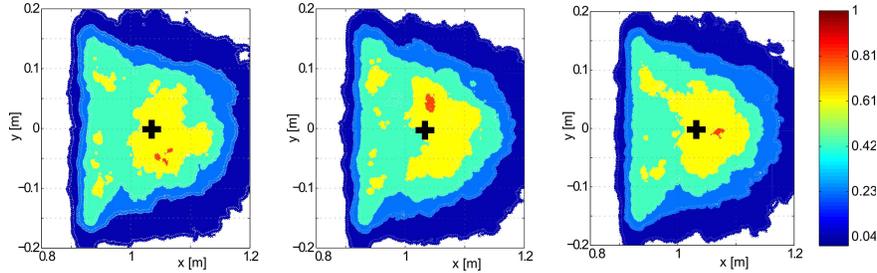}
  \caption{The discharge flow through a horizontal cross-section beneath
  the end of the conveyor.  The contour plots show the accumulated particles distribution 
 from the reference simulations $\textsub{N}{it} = 500$ (left), $\textsub{N}{it} = 150$ (right)
 and simulation with model reduction using contact and sensor splitting (right).  The crosses mark
 the geometric centre of the particle distributions.}
  \label{fig:discharge}
\end{figure}

The resting angle of repose of the pile formed on the conveying surface is found to be $40.3\pm0.8^{\circ}$
using contact splitting.  This is in good agreement with the references $41.5\pm0.7^{\circ}$ for $\textsub{N}{it} = 500$ and $40.6\pm0.5^{\circ}$ for $\textsub{N}{it} = 150$.  The discharge flow at the end of the conveyor is presented in 
Fig.~\ref{fig:discharge}. The cross-sectional flow distribution is very similar in the three simulations.  
The model order reduction level was steady around $h = 85$ \% during the simulation.

\begin{figure}
\centering
  \includegraphics[width=0.80\textwidth,]{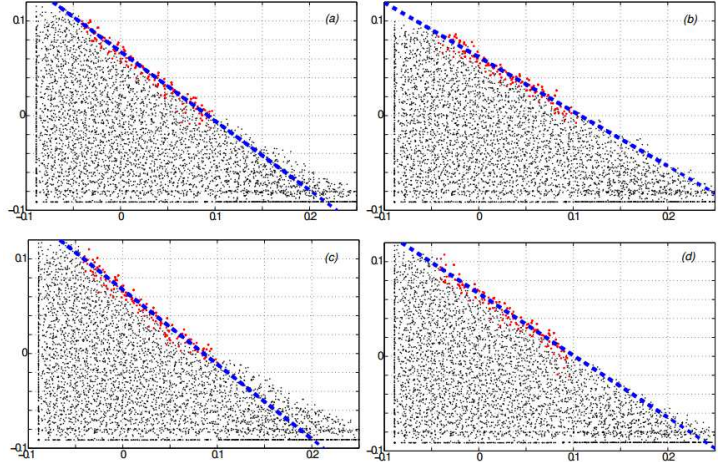}
  \caption{The final particle distribution in the granular collapse simulations: 
  (a) $\textsub{N}{it} = 500$ reference; (b) $\textsub{N}{it} = 150$ reference;
  (c) model reduction tr-15-50-f; (d) model reduction tr-15-50-fv.
  The blue line indicate the angle of repose computed from the surface defined
  by the red particles.}
  \label{fig:dam_break_profiles}
\end{figure}

\begin{figure}
\centering
  \includegraphics[width=0.85\textwidth, trim = 0 0 0 0, clip]{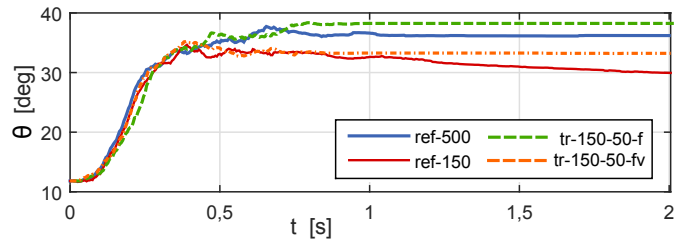}
  \caption{Evolution of the top surface angle during granular collapse.}
  \label{fig:dam_break_angle}
\end{figure}

\begin{figure}
\centering
  \includegraphics[width=0.75\textwidth, trim = 0 0 0 0, clip]{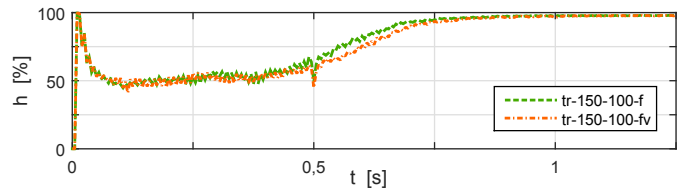}
  \caption{Model reduction level as function of time in granular collapse.}
  \label{fig:dam_break_reduction_level}
\end{figure}

The distribution of particles after the granular collapse are displayed
in Fig.~\ref{fig:dam_break_profiles}.  The positions are measured at time $t = 2$ s.
The angle of repose, $\textsub{\theta}{collapse}$, of the $\textsub{N}{it} = 500$  reference pile is
$36.2^{\circ}$, to be compared with $29.9^{\circ}$ for the $\textsub{N}{it} = 150$
reference and $37.5^{\circ}$ and $34^{\circ}$ for model reduction with background trial solve with force (f) and force and velocity (fv) split conditions, respectively.
The evolution of the inclination angle of the top surface of 
the collapsing cube is found in Fig.~\ref{fig:dam_break_angle}.
The initial collapse is similar in all the simulations except the
\emph{tr-150-50-f} that evolve somewhat slower initially.
The $\textsub{N}{it} = 150$ reference
simulation reach as highest $34^{\circ}$ and then decrease gradually due to insufficient
sliding and rolling resistance with that number of iterations.  The model reduction simulation
become rigid at its maximum angle and fail to resolve the final relaxation of the slope
through small avalanches, that are present in the $\textsub{N}{it} = 500$ reference simulation.
The combination of contact split and background trial split did not resolve this.
The evolution of the model reduction level is found in Fig.~\ref{fig:dam_break_reduction_level}.
In the background trial solve split simulation the reduction level varies between $50$ and $100$ \%.
This is close to the theoretical maximum for the given thresholds, found by analysing the $\textsub{N}{it} = 500$ reference
simulation.

\begin{figure}
\centering
  \includegraphics[width=0.85\textwidth, trim = 0 0 0 0, clip]{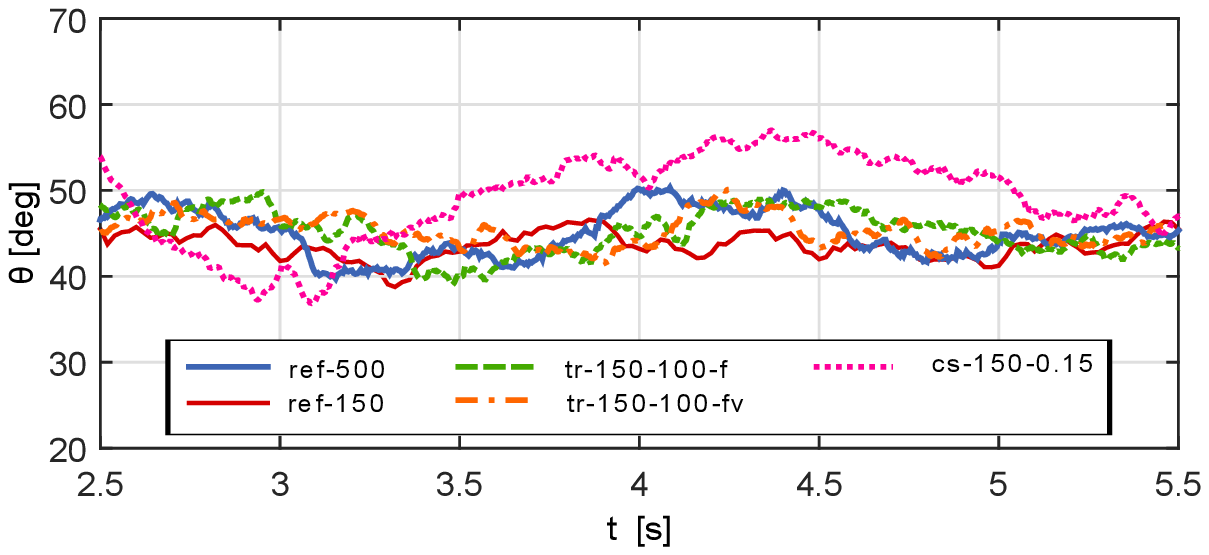}
  \caption{Evolution of the dynamic angle of repose in the rotating drum.}
  \label{fig:drum_angle_evolution}
\end{figure}

\begin{figure}
\centering
  \includegraphics[width=0.85\textwidth, trim = 0 00 0 00, clip]{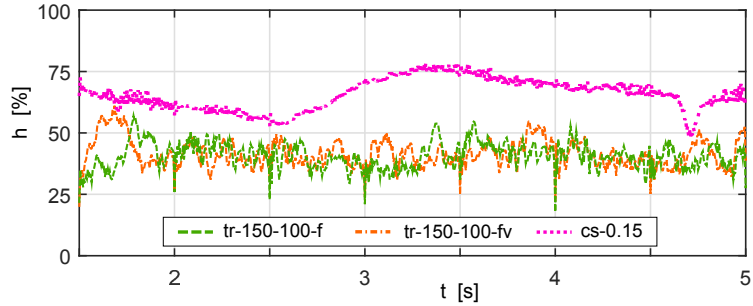}
  \caption{Model reduction level $h$ as function of time in the rotating drum. }
  \label{fig:drum_reduction_level}
\end{figure}

Sample states from model reduction of drum flow simulations are presented in 
Fig.~\ref{fig:drum_flow}.  The evolution of the dynamic angle of repose
is found in Fig.~\ref{fig:drum_angle_evolution}.  
The time-averaged angle of repose for the reference 
$\textsub{N}{it} = 500$ is $\textsub{\theta}{drum} = 44\pm3^{\circ}$. 
The background trial solve split method produces a flow with angle $44\pm3^{\circ}$ 
while the contact plus sensor split methods produces a flow with angle $49\pm5^{\circ}$
and with notable artefacts appearing as structures with angle much larger than the angle 
of repose and high lifting of material in the drum.
This is also the reason for the bigger variation on the averaged angle of repose.
No thresholds were found for the contact split method alone
that showed any significant model reduction but did not produce large approximation errors (overly rigid).
No thresholds were found for the contact split method alone
that showed any significant model reduction but did not produce large approximation errors (overly rigid).
The model reduction level over time is presented in Fig.~\ref{fig:drum_reduction_level}.  
For the contact method plus sensor split method it varies between  $50\--75$ \%.  The
background trial solve split oscillate between $25$ and $50$\%.  No parameters were found that gave
higher level of reduction without increase of artifacts in the dynamics. 

\section{Computational acceleration}
The potential computational acceleration of using model order reduction in
NDEM simulations is estimated and discussed in this section.
The time required for simulating $\textsub{t}{real}$ seconds of evolution 
is the product of the number of time steps and the computational time for each step
\begin{equation}
	\textsub{t}{comp} = \frac{\textsub{t}{real}}{\dt} \cdot 
	\left[ \textsub{t}{coll} + \textsub{t}{mod} + \textsub{t}{solve} \right] 
\end{equation}
with time serially separated in collision detection, $\textsub{t}{coll}$, 
model order reduction, $\textsub{t}{mod}$ and solver time, $\textsub{t}{solve}$.  
We define the computational speed-up from model order reduction as
$S \equiv t^0_\text{comp} / \textsub{t}{comp}(h)$, where $t^0_\text{comp}$ refer to 
a simulation of the fully resolved system without adaptive model reduction, while $\textsub{t}{comp}(h)$
is the time for a simulation with model reduction level $h$.
It is characteristic for NDEM simulations that $\textsub{t}{solve}\gg \textsub{t}{coll}$,
e.g., 88\% of the total time was reported in \cite{renouf:2004:pvn}.  We therefore discard collision detection time from here on.  When using
PGS, the solve time can be estimated by $\textsub{t}{solve} = \textsub{K}{cpu} \cdot \textsub{N}{c} \cdot \textsub{N}{it} / S_{\parallel}$, where $\textsub{K}{cpu}$ is the computational time for doing a single 
contact constraint solve, $\textsub{N}{c} \sim \textsub{n}{p}\textsub{\tilde{N}}{p}$ is the number of 
contact constraints assuming on average $2\textsub{n}{p}$ contacts per particle, and 
$S_{\parallel}(\textsub{N}{cpu})$ is the parallel speed-up of $\textsub{N}{cpu}$ cores.  Given a spatial error 
tolerance, $\epsilon$, the required number of iterations scale with the number of particles as 
$\textsub{N}{it} = c \textsub{\tilde{N}}{p}^{\gamma}/\epsilon$ \cite{servin:2014:esn},
for constant $c\approx 0.1$ and exponent $\gamma = 1/\textsub{n}{D}$ that depend
on the whether the system is close to a linear column ($\textsub{n}{D} = 1$) , 2D plane 
($\textsub{n}{D} = 2$) or a 3D volumetric system ($\textsub{n}{D} = 3$).  The solve time can thus be written
\begin{equation}
	\textsub{t}{solve}(h) = c\textsub{K}{cpu}\left[ (1 - h) N_\text{p}\right]^{1+\gamma}/\epsilon S_{\parallel}.
\end{equation}

The computational overhead for doing merge and refinement
using contact or sensor based splitting do not involve more than one pass through the
contact network.
The computational time can thus be estimated by 
$\textsub{t}{mod} = \alpha \textsub{K}{cpu}\textsub{n}{p} \textsub{N}{p}/S_{\parallel}$ for some 
constant $\alpha < 1$, since the operations do not involve solving the local contact problem.  This imply the following speed-up
\begin{equation}
	S  \approx  \frac{1}{\alpha \textsub{N}{p}^ {-\gamma} + (1 -h)^{1 + \gamma}}.
\end{equation}

Using background trial solve split for model reduction is more demanding.  If a PGS background
solve can be limited to fraction $\beta< 1$ of the full system and run with a larger error tolerance,
 $\epsilon_{\text{mod}} > \epsilon$, the computational overhead can be estimated to
 $\textsub{t}{mod}  = c\textsub{K}{cpu}\left[ \beta N_\text{p}\right]^{1+\gamma}/\epsilon_{\text{mod}} S_{\parallel}$
 and the speed-up become 
\begin{equation}
	S \approx	\frac{ 1 }{  \beta^{1 + \gamma} \cdot  \epsilon/\epsilon_{\text{mod}}  + (1 -  h)^{1 + \gamma}  }.
\end{equation}

\begin{figure}
\centering
  \includegraphics[width=0.8\textwidth, trim = 0 0 0 0, clip]{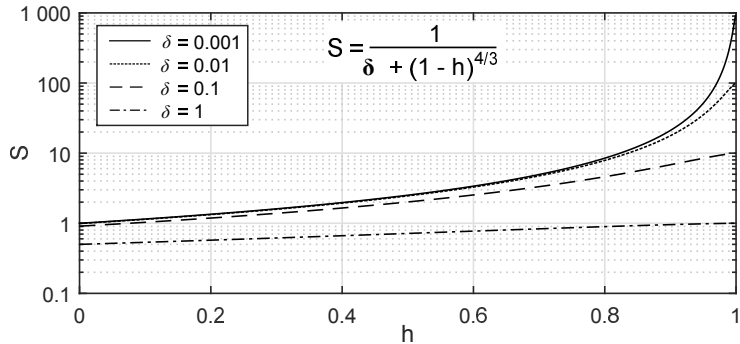}
  \caption{The computational speed-up depending on model reduction level and
  computational overhead.}
  \label{fig:speedup}
\end{figure}

The computational speed-up, depending on model reduction level $h$ and overhead
factor $\delta = \alpha \textsub{N}{p}^ {-\gamma}$ or 
 $\delta = \beta^{1 + \gamma} \cdot  \epsilon/\epsilon_{\text{mod}}$,
is plotted in Fig.~\ref{fig:speedup} for the case of 3D volumetric systems ($\gamma = 1/3$).
A significant speed-up of a factor $5$ can be achived
even at the modest model reduction level $h = 0.7$.  The speed-up can reach
up to $50$ at $h = 0.95$ and overhead $\delta = 0.01$.  The potential speed-up is lost almost entirely if the computational overhead cannot 
be made lower than $\delta = 0.1$.  This should be easy to accomplish in the case of contact and
sensor based splitting, especially for large systems where $\delta \propto \textsub{N}{p}^ {-\gamma}$.
Achieving a significant speed-up from background trial solve split thus relies on 
$\beta^{1 + \gamma} \cdot  \epsilon/\epsilon_{\text{mod}} \lesssim 0.1$.  This
can be reached if it is sufficient to apply background solve on a fraction of $\beta = 0.2$
of the full system or use a background error tolerance $\epsilon_{\text{mod}} = 10 \cdot \epsilon$.
Combining the two possibilities, an overhead of about $0.1$ can be reached by
$\beta = 0.5$ and $\epsilon_{\text{mod}} = 4 \cdot \epsilon$.
The prototype implementation do not scale sufficiently well for large systems to verify these 
estimates.  To ease prototyping and experimentation with different algorithms and parameters
for model order reduction these routines were implemented in a Lua scripting framework 
which lead to an unnecessary amount of data copying of states. 


\section{Conclusion and discussion}
\label{sec:discussion}

A method for adaptive model order reduction for nonsmooth discrete element 
simulation has been developed and analysed.  In the reduced model 
rigid aggregate bodies are substituted for collections of contacting particles 
collectively moving as rigid bodies.  Conditions for model reduction and refinement
are derived from a model approximation error.  The scaling analysis show that the
computational performance may be increased by $5 - 50$ times for a model
reduction level between $70 - 95$ \% given that the computational overhead do not exceed
the given scaling conditions.  The method is highly applicable for granular systems  with large
regions in resting or rigidly co-moving state over long periods.  It is less efficient and harder to parametrize
for systems with sudden and frequent transitions between rigid to liquid or gaseous regime
and it is directly inappropriate for systems dominated by shear motion.
Furthermore, the presented method is fully compatible with rigid multibody dynamics and can 
support particles merging with articulated  mechanisms, such as the excavator in
Fig.~\ref{fig:illustration}.

A number of observations were made in the numerical experiments.  When doing
model refinement based on contact events, it is in general insufficient to 
split only particles that are impacted directly.  The refinement typically need to 
propagate further into the contact network.  The refinement depth  $N^\text{spl}_{c} = 2$ was
used in the experiments.  Refinement based on contact separation events was found to be particularly 
difficult to parametrize and was eventually not used at all.  Small changes in the separation threshold 
easily altered the behaviour from non-responding to splitting 
propagating throughout the contact network.  As an effect, contact event based refinement
is not reliable for simulating quasi-stable configurations and gravity driven flow.  In the
absence of impacts, rigid aggregates remain rigid indefinitely.   Split sensors can be used to remedy
this when there is knowledge on where to place such sensors to guarantee refinement.

When using background trial solve it was found that the 
force balance condition alone give reliable prediction for 
model refinement.  No improvement was found by adding torque balance or contact velocity conditions.  Using background solve, velocity condition alone led to higher fluctuations in the model order reduction level than using both force and velocity conditions.  

To increase the applicability of adaptive model order reduction for discrete element
methods further, the reduced model need to be extended beyond rigid aggregates to 
elastic and shearing modes.  The application of model order reduction to conventional 
smooth DEM is also an interesting question to address.

\section*{Appendix}
\label{sec:appendix}

\subsection*{A. Numerical integration of NDEM}
\label{sec:appendix_A}
The numerical time integration scheme is based on the \textsc{SPOOK} stepper \cite{lacoursiere:2007:rvs}
derived from discrete variational principle for the augmented system 
$(\tv{x}, \tv{v}, \bg{\lambda}, \dot{\bg{\lambda}})$ and applying a
semi-implicit discretization.  The stepper is linearly stable and $\mathcal{O}(\dt^2)$ accurate 
for constraint violations \cite{lacoursiere:2007:rvs}.  Stepping 
the system position and velocity, $(\tv{x}_i, \tv{v}_i) \to (\tv{x}_{i+1}, \tv{v}_{i+1})$,
from time $t_i$ to $t_{i+1} = t_i + \dt$ involve solving a 
mixed complementarity problem (MCP) \cite{murty:1988:lcl}.  For a NDEM system the
MCP take the following form

\begin{equation}
	  \label{eq:MCP}
		\begin{array}
	      		[c]{c}%
				\tm{H}\tv{z}  + \tv{b}  = \tv{w}_{l} - \tv{w}_{u} \\
				\mbox{\scriptsize $ 0 \leq \tv{z} - \tv{l} \perp \tv{w}_{l}\geq 0$ } \\
				\mbox{\scriptsize $ 0 \leq \tv{u} - \tv{z} \perp \tv{w}_{u}\geq 0$ }
	\end{array}
\end{equation}
where
\begin{equation}
	  \label{eq:H}	
	\tm{H} = \left[
		\begin{array}
	      		[c]{cccc}%
	      		\tm{M} & - \textsub{\tm{G}}{n}^\T & -\textsub{\tm{G}}{t}^\T & -\textsub{\tm{G}}{r}^\T\\
	      		\textsub{\tm{G}}{n} & \textsub{\bg{\Sigma}}{n} & 0 & 0 \\
			\textsub{\tm{G}}{t} & 0 & \textsub{\bg{\Sigma}}{t} & 0\\
			\textsub{\tm{G}}{r} & 0 & 0 & \textsub{\bg{\Sigma}}{r} \\
	    \end{array}
		\right],
\end{equation}
\begin{equation}\label{eq:zb}
		\tv{z} = \begin{bmatrix} {\tv{v}_{i+1}} \\ \bg{\lambda}_{\text{n},i+1} \\ \bg{\lambda}_{\text{t},i+1} \\ \bg{\lambda}_{\text{r},i+1}\end{bmatrix},\ \
		\tv{b} = \begin{bmatrix} {-\tm{M} \tv{v}_{i} -\dt\tm{M}^{-1} \textsub{\tv{f}}{ext} } \\ \tfrac{4}{\dt} \textsub{\bg{\Upsilon}}{n}\textsub{\tv{g}}{n} - \textsub{\bg{\Upsilon}}{n}\textsub{\tm{G}}{n}\tv{v}_{i} \\ {0} \\ {0}\end{bmatrix},
\end{equation}
%
and the solution vector $\tm{z}$ contains the new velocities and the
Lagrange multipliers $\bg{\lambda}_{\text{n}}$, $\bg{\lambda}_{\text{t}}$ and $\bg{\lambda}_{\text{r}}$.
For notational convenience, a factor $\dt$ has been absorbed in 
the multipliers such that the constraint force reads $\tm{G}^T\bg{\lambda}/\dt$.
The upper and lower limits, $u$ and $l$ in Eq.~(\ref{eq:MCP}), 
follow from Signorini-Coulomb and rolling resistance law 
with the friction and rolling resistance coefficients $\textsub{\mu}{t}$ and $\textsub{\mu}{r}$.
Since the limits depend on the solution this is a partially nonlinear
complementarity problem.  $w_{l}$ and $w_{u}$ are temporary slack variables.
Each contact $n$ between body $a$ and $b$ add contributions to the constraint vector and
normal, friction and rolling Jacobians according to
\begin{eqnarray}
\label{eq:constraint}
g_{(n)} & = &\vec{\tv{n}}_{(n)}^\T(\vec{\tv{x}}_{[a]} + \vec{\tv{d}}^{(n)}_{[a]} -\vec{\tv{x}}_{[b]} - \vec{\tv{d}}^{(n)}_{[b]})^{\textsub{e}{H}}_{(\alpha)}\nonumber,\\
\tm{G}_{\text{n}[a]}^{(n)} & = & \textsub{e}{H} g^{\textsub{e}{H}-1}_{(\alpha) }\left[ 
	\begin{array}
	      		[c]{cc}%
	      		-\vec{\tv{n}}_{(n)}^{\T} &\ \  -(\vec{\tv{d}}^{(n)}_{[a]} \times\vec{\tv{n}}_{(n)})\tran{T}	    	
	\end{array} \right]\nonumber,\\
\tm{G}_{\text{n}[b]}^{(n)} & = & \textsub{e}{H} g^{\textsub{e}{H}-1}_{(\alpha) }\left[ 
	\begin{array}
	      		[c]{cc}%
	      		\vec{\tv{n}}_{(n)}^{\T} &\ \  (\vec{\tv{d}}^{(n)}_{[b]} \times\vec{\tv{n}}_{(n)})\tran{T}
	\end{array} \right],\\
\tm{G}_{\text{t}[a]}^{(n)} & = & \left[ 
	\begin{array}
	      		[c]{cc}%
	      		-\vec{\tv{t}}^{(n)\T}_1 &\ \  -(\vec{\tv{d}}^{(n)}_{[a]} \times \vec{\tv{t}}^{(n)}_1)\tran{T}	    	\\
	      		-\vec{\tv{t}}^{(n)\T}_2 &\ \  -(\vec{\tv{d}}^{(n)}_{[a]} \times\vec{\tv{t}}^{(n)}_2)\tran{T}
	      		\end{array} \right]\nonumber,\\
\tm{G}_{\text{t}[b]}^{(n)} & = & \left[ 
	\begin{array}
	      		[c]{cc}%
	      		\vec{\tv{t}}^{(n)\T}_1 &\ \  (\vec{\tv{d}}^{(n)}_{[b]} \times\vec{\tv{t}}^{(n)}_1)\tran{T}	    	\\
	      	 	\vec{\tv{t}}^{(n)\T}_2 &\ \  (\vec{\tv{d}}^{(n)}_{[b]} \times\vec{\tv{t}}^{(n)}_2)\tran{T}
	      		\end{array} \right]\nonumber,\\
	\label{eq:Jacobian_rolling}
	\tm{G}_{\text{r}[a]}^{(n)} & = &
	\left[
	\begin{array}
	      		[c]{cccc}%
				\mathbf{0}_{1\times 3} &\ \vec{\tv{t}}^{(n)\T}_1 &\ \  \mathbf{0}_{1\times 3} & -\vec{\tv{t}}^{(n)\T}_1\\
				\mathbf{0}_{1\times 3} &\ \vec{\tv{t}}^{(n)\T}_2 &\ \  \mathbf{0}_{1\times 3} & -\vec{\tv{t}}^{(n)\T}_2\\
				\mathbf{0}_{1\times 3} &\ \vec{\tv{n}}^{(n)\T} &\ \  \mathbf{0}_{1\times 3} & -\vec{\tv{n}}^{(n)\T}\\
	\end{array} \right]\nonumber,\\
	\tm{G}_{\text{r}[b]}^{(n)} & = &
	\left[
	\begin{array}
	      		[c]{cccc}%
				\mathbf{0}_{1\times 3} &\ \ -\vec{\tv{t}}^{(n)\T}_1 &\ \  \mathbf{0}_{1\times 3} &\vec{\tv{t}}^{(n)\T}_1\\
				\mathbf{0}_{1\times 3} &\ \ -\vec{\tv{t}}^{(n)\T}_2 &\ \  \mathbf{0}_{1\times 3} &\vec{\tv{t}}^{(n)\T}_2\\
				\mathbf{0}_{1\times 3} &\ \ -\tv{n}^{(n)\T} &\ \  \mathbf{0}_{1\times 3} &\vec{\tv{n}}^{(n)\T}\\
	\end{array} \right]\nonumber,
\end{eqnarray}
where $\vec{\tv{d}}^{(n)}_{[a]}$ and $\vec{\tv{d}}^{(n)}_{[b]}$ are the positions of the contact point $n$ relative to the particle positions $\vec{\tv{x}}_{[a]}$ and $\vec{\tv{x}}_{[b]}$.  The orthonormal contact normal and tangent vectors are $\vec{\tv{n}}_{(n)}$, $\vec{\tv{t}}_{(n)_1}$ and $\vec{\tv{t}}_{(n)_2}$.  For linear contact model 
$\textsub{e}{H} = 1$ and for
the nonlinear Hertz-Mindlin model $\textsub{e}{H} = 5/4$.  The diagonal matrices 
$\textsub{\bg{\Sigma}}{n}$, $\textsub{\bg{\Sigma}}{t}$ , $\textsub{\bg{\Sigma}}{r}$
and $\textsub{\bg{\Upsilon}}{n}$ contain the contact material parameters and
are as follows
\begin{eqnarray}
	  \label{eq:diagonal}
	\textsub{\bg{\Sigma}}{n} & = & \frac{4}{\Delta t^2}\frac{\textsub{\varepsilon}{n}}{1+4\tfrac{\textsub{\tau}{n}}{\Delta t}}\bm{1}_{N_{c}\times N_{c}}\nonumber,\\ 
	\textsub{\bg{\Sigma}}{t} & = & \frac{\textsub{\gamma}{t}}{\Delta t}\bm{1}_{2N_{c}\times 2N_{c}}\nonumber,\\
	\textsub{\bg{\Sigma}}{r} & = & \frac{\textsub{\gamma}{r}}{\Delta t}\bm{1}_{3N_{c}\times 3N_{c}},\\
	\textsub{\bg{\Upsilon}}{n} & = & \frac{1}{1+4\tfrac{\textsub{\tau}{n}}{\Delta t}}\bm{1}_{N_{c}\times N_{c}}\nonumber.
\end{eqnarray}
The MCP parameters map to DEM material parameters by $\textsub{\varepsilon}{n} = \textsub{e}{\tiny H}/\textsub{k}{n}$, 
$\textsub{\gamma}{n}^{-1} = \textsub{k}{n}c/e^2_{\text{\tiny H}}$ and $\textsub{\tau}{n} = \max(\textsub{n}{s}\Delta t, \textsub{\varepsilon}{n}/\textsub{\gamma}{n})$, with elastic stiffness coefficient $\textsub{k}{n}$
and viscosity $c$.  For the Hertz-Mindlin contact law, 
$\textsub{k}{n} = \textsub{e}{\tiny H}E\sqrt{r^*}/3(1-\nu^2)$
where $r^{*} = (r_a + r_b)/r_a r_b$ is the effective radius, $E$ is the Young\rq{}s modulus and 
$\nu$ is the Poisson ratio.   For small relative contact 
velocities the normal force approximates $\tm{G}_{\text{n}}^{(n)\T}\bg{\lambda}_{\text{n}}^{(n)} \approx  \varepsilon_{\text{n}}^{-1}\tm{G}_{\text{n}}^{(n)\T}\tv{g}_{\text{n}}^{(n)} = 
\pm \textsub{k}{n}\left[ \textsub{\tm{g}}{n}^{2\textsub{e}{\tiny H} -1} + c \textsub{\tm{g}}{n}^{2(\textsub{e}{\tiny H} -1)}\dot{g}_{\text{n}}\right]\bm{n}$.  
High ingoing velocities are treated as impacts and this is done 
\emph{post facto}.  After stepping the velocities and 
positions an impact stage follows.  This include solving a 
MCP similar to Eq.~(\ref{eq:MCP}) but with the Newton impact law,
$\textsub{\tm{G}}{n}^{(n)} \tv{v}_{+} = - e\textsub{\tm{G}}{n}^{(n)} \tv{v}_{-}$,
replacing the normal constraints for the contacts with normal
velocity larger than an impact velocity threshold $\textsub{v}{imp}$.
The remaining constraints are maintained by imposing $\tm{G}^{(n)}\tv{v}_{+} = 0$.
This can be expressed by a matrix multiplication $\textsub{\tm{G}}{n} \tv{v}_{+} = - \tv{E}\textsub{\tm{G}}{n}\tv{v}_{-}$, where the diagonal matrix $\tv{E}$ values alternate
between $e$ and $0$ for impacting and resting contacts, respectively.
The MCP is solved using a projected Gauss-Seidel (PGS) algorithm, as described in 
Ref.~\cite{servin:2014:esn}.  The algorithm is listed in Algorithm \ref{alg:NDEMPGS}.
The NDEM method with PGS solver is implemented in the software AgX Dynamics \cite{agx}.  
In the present study $\textsub{\gamma}{t} = \textsub{\gamma}{r} = 10^{-6}$, $\textsub{n}{s} = 2$
was used and a linear contact model, $\textsub{e}{\tiny H} = 1$,  for consistency with contacts 
between elementary rigid bodies and aggregate bodies, for which linear contact constraints
are default in AgX.  

\begin{algorithm}[h!]
  \caption{PGS solver for the MCP}\label{alg:NDEMPGS}
  \begin{algorithmic}
    \If{impact stage}
     		\State $\textsub{\tv{b}}{n} = \tv{E} \textsub{\tv{G}}{n}\tv{v}$
     \ElsIf{continuous stage}
    		\State $\textsub{\tv{b}}{n} = (4/\Delta t)
 \textsub{\bg{\Upsilon}}{n}\textsub{\tv{g}}{n} - \textsub{\bg{\Upsilon}}{n}\textsub{\tv{G}}{n}\tv{v}$
 		  \State pre-step $\tv{v} = \tv{v} + \Delta t \tv{M}^{-1} \textsub{\tv{f}}{ext}$
 	\EndIf
  \State $\tv{q} = [-\textsub{\tv{b}}{n}^T, 0,0]^T$ 
  \For{$k = 1,\hdots,\textsub{N}{it}$ and {\bf while} \emph{criteria}$(\tv{r})$} 
  \For{each contact $n = 0,1,\hdots,\textsub{N}{c}-1$}
  \For{each constraint $\alpha$ of contact $n$}
	  \State $\tv{r}^{(n)}_{\alpha,k} = - \tv{q}^{(n)}_{\alpha,k} + \tv{G}^{(n)}_{\alpha}\tv{v}$ \Comment{residual}
	  \State $\bg{\lambda}^{(n)}_{\alpha,k} = \bg{\lambda}^{(n)}_{\alpha,k-1} + \tv{D}^{-1}_{\alpha,(n)} \tv{r}^{(n)}_{\alpha,k}$ \Comment{multiplier}
	  \State $\bg{\lambda}^{(n)}_{\alpha,k} \leftarrow \text{proj}_{\mathcal{C}_{\mu}}(\bg{\lambda}^{(n)}_{k})$ \Comment{project}
	  \State $\Delta \bg{\lambda}^{(n)}_{\alpha,k} =  \bg{\lambda}^{(n)}_{\alpha,k} - \bg{\lambda}^{(n)}_{\alpha,k-1}$
	  \State $\tv{v}  =  \tv{v} + \tv{M}^{-1}\tv{G}^{T}_{\alpha,(n)}\Delta\bg{\lambda}^{(n)}_{\alpha,k}$
  \EndFor
  \EndFor
  \EndFor
  \end{algorithmic}
\end{algorithm}

\section*{Acknowledgment}
\label{sec:acknowledgments}
This project was supported by Algoryx Simulations, LKAB (dnr 223-2442-09), 
UMIT Research Lab and VINNOVA (2014-01901).

\def\cprime{$'$}


\begin{thebibliography}{10}
\expandafter\ifx\csname url\endcsname\relax
  \def\url#1{\texttt{#1}}\fi
\expandafter\ifx\csname urlprefix\endcsname\relax\def\urlprefix{URL }\fi
\expandafter\ifx\csname href\endcsname\relax
  \def\href#1#2{#2} \def\path#1{#1}\fi

\bibitem{poeschel:2005:cgd}
T.~P{\"{o}}schel, T.~Schwager, Computational Granular Dynamics, Models and
  Algorithms, Springer-Verlag, 2005.

\bibitem{Radjai:2009:cdn}
F.~Radjai, V.~Richefeu, Contact dynamics as a nonsmooth discrete element
  method, Mechanics of Materials 41~(6) (2009) 715--728.

\bibitem{Moreau:1999:NAS}
J.~J. Moreau, Numerical aspects of the sweeping process, Computer Methods in
  Applied Mechanics and Engineering 177 (1999) 329--349.

\bibitem{Jean:1999:NSC}
M.~Jean, The non-smooth contact dynamics method, Computer Methods in Applied
  Mechanics and Engineering 177 (1999) 235--257.

\bibitem{Antoulas:2005:als}
A.~Antoulas, Approximation of Large-Scale Dynamical Systems,
Society for Industrial and Applied Mathematics (2005).

\bibitem{Kerschen:2005:mpo}
G.~Kerschen,  J-C.~Golinval, A.~Vakakis, L.~Bergman,
The Method of Proper Orthogonal Decomposition for Dynamical
Characterization and Order Reduction of Mechanical Systems:
An Overview, Nonlinear Dynamics 41(2005) 147--169.

\bibitem{Nowakowski:2012:mor}
C.~Nowakowski, J.~ Fehr, M.~Fischer, P.~Eberhard,
Model Order Reduction in Elastic Multibody Systems using the Floating Frame of Reference Formulation,
In Proceedings MATHMOD 2012-7th Vienna International Conference on Mathematical Modelling, Vienna, Austria, (2012).

\bibitem{Glossman:2010:rde}
P.~Gl{\"o}smann, Reduction of discrete element models by Karhunen–Loève transform: a hybrid model approach,
Computational Mechanics 45~(4) (2010)  375--385.

\bibitem{Munjiza:2004:cfd}
A.~Munjiza, The Combined Finite-Discrete Element Method, Wiley, (2004).

\bibitem{Miehe:2010:hts}
C.~Miehe∗, J.~Dettmar and D. Zah, 
Homogenization and two-scale simulations of granular materials for different microstructural constraints, Int. J. Numer. Meth. Engng 83 (2010) 1206–-1236.

\bibitem{Guo:2014:cfd}
N.~Guo, J.~Zhao, A coupled FEM/DEM approach for hierarchical multiscale modelling of granular media, Int. J. Numer. Meth. Engng 99~(11) (2014) 789--818.

\bibitem{Wellman:2012:cfd}
C.~Wellmann, P.~ Wriggers, A two-scale model of granular materials, Computer Methods in Applied Mechanics and Engineering, 205-208 (2012), 46--58.

\bibitem{Redon:2005:eea}
S.~Redon, M.~Lin, An Efficient, Error-Bounded Approximation Algorithm for Simulating Quasi-Statics of Complex Linkages,
In Proceedings of ACM Symposium on Solid and Physical Modeling (2005).

\bibitem{Redon:2005:ada}
S.~Redon, N.~Galoppo,M.~Lin,
Adaptive Dynamics of Articulated Bodies,
In ACM Transactions on Graphics (SIGGRAPH 2005), 24~(3) (2015).

\bibitem{Cho:2007:rcpm}
N.~Cho, C.~D.~Martin, D.~C.~Sego, A clumped particle model for rock,
International Journal of Rock Mechanics \& Mining Sciences 44 (2007) 997--1010.

\bibitem{Jaeger:1996:gsl}
H.~Jaeger, S.~Nagel, R.~Behringer, Granular solids, liquids, and gases,
Rev. Mod. Phys. 68~(4) (1996) 1259--1273.

\bibitem{bornemann:1997:hhs}
Bornemann~F.,Sch{\"{u}}tte~C.
\newblock Homogenization of {Hamiltonian} systems with a strong constraining
  potential.
\newblock {\em Phys. D}, 102(1-2):57--77, 1997.

\bibitem{lacoursiere:2007:rvs}
C.~Lacoursi{\`{e}}re, Regularized, stabilized, variational methods for
  multibodies, in: D.~F. Peter~Bunus, C.~F{\"{u}}hrer (Eds.), The 48th
  Scandinavian Conference on Simulation and Modeling (SIMS 2007), 
  Link{\"{o}}ping University Electronic Press, 2007, pp. 40--48.

\bibitem{servin:2014:esn}
M.~Servin, D.~Wang, C.~Lacoursi{\`e}re, K.~Bodin, Examining the smooth and
  nonsmooth discrete element approach to granular matter, Int. J. Numer. Meth.
  Engng. 97 (2014) 878--902.

\bibitem{renouf:2005:nst}
M.~Renouf, D.~Bonamy, F.~Dubois, P.~Alart, Numerical simulation of two-dimensional steady granular flows in rotating drum: On surface flow rheology, Phys.~Fluids 17 (2005).  

\bibitem{agx}
Algoryx Simulations, AgX Dynamics User Guide Version 2.12.1.0, December 2014.

\bibitem{lua}
{Lua}, \url{http://www.lua.org}, October 2015.

\bibitem{renouf:2004:pvn}
M.~Renouf, F.~Dubois P.~Alart, A parallel version of the non smooth contact dynamics algorithm applied to the simulation of granular media, Journal of Computational and Applied Mathematics 168(1-2), 2004, pp. 375-–382.

\bibitem{murty:1988:lcl}
K.~G. Murty, Linear Complementarity, Linear and Nonlinear Programming,
  Helderman-Verlag, Heidelberg, 1988.

\end{thebibliography}
\end{document}